\documentclass[oldversion,PRINTER]{aa}  
\usepackage{graphicx}
\usepackage{rotating}
\usepackage{txfonts}
\begin{document}
\newcommand{\bm}[1]{\mbox{\boldmath$#1$\unboldmath}}
\newcommand{\bmi}[1]{\mbox{\boldmath$\scriptstyle{#1}$\unboldmath}}
\newcommand{\varcsec}{^{\prime\prime}}
\renewcommand{\arcsec}{.\hspace{-0.9mm}'\!\hskip0.4pt'\hspace{-0.2mm}}
   \title{High-resolution spectro-polarimetry of a flaring sunspot penumbra}

%   \subtitle{I. xxx}

   \author{J. Hirzberger\inst{1}
          \and
          T. Riethm\" uller\inst{1}
          \and
          A. Lagg\inst{1}
          \and
          S.K. Solanki\inst{1,2}
          \and
          P. Kobel\inst{1}
          }

   \offprints{J. Hirzberger}

   \institute{Max-Planck-Institut f\" ur Sonnensystemforschung, 
              Katlenburg-Lindau, Germany\\
              \email{hirzberger@mps.mpg.de}
   \and
              School of Space Research, Kyung Hee University, Yongin,
              Gyeonggi 446-71, Korea}

   \date{Received: ?? 2009; accepted June 4, 2009}

% \abstract{}{}{}{}{} 
% 5 {} token are mandatory
 
  \abstract{We present simultaneous photospheric and chromospheric
  observations of the trailing sunspot in NOAA~10904 during a weak
  flare eruption (GOES magnitude B7.8), obtained with the Swedish
  Solar Telescope (SST) in La Palma, Canary Islands. High-resolution
  \ion{Ca}{ii}\,$H$ images show a typical two-ribbon structure that
  has been hitherto only known for larger flares, and the flare
  appears in a confined region that is discernible by a bright
  border. The underlying photosphere shows a disturbed penumbral
  structure with intersecting branches of penumbral
  filaments. High-resolution Doppler- and vector-magnetograms exhibit
  oppositely directed Evershed flows and magnetic field vectors in the
  individual penumbral branches, resulting in several regions of
  magnetic azimuth discontinuity and several islands where the
  vertical magnetic field is reversed. The discontinuity regions are
  co-spatial with the locations of the onset of the flare
  ribbons. From the results, we conclude that the confined flare
  region is detached from the global magnetic field structure by a
  separatrix marked by the bright border visible in
  \ion{Ca}{ii}\,$H$. We further conclude that the islands of reversed
  vertical field appear because of flux emergence and that the strong
  magnetic shear appearing in the regions of magnetic azimuth
  discontinuity triggers the flare. 
% {a} 
% {b} 
% {c} 
% {d} 
% {e} 
% 
} \maketitle
%
%----------------------------------------------------------------
%

\section{Introduction}

Solar flares are most likely caused by a sudden release of magnetic
energy due to plasma instabilities or magnetic reconnection. The
dynamics of large X-class flares has been intensely studied in the
past decades (see e.g. Li et al. 2000a,b; Wang 2006; Su et al. 2007,
Benz 2008 and references therein). The results are still inconclusive
for the trigger, although magnetic shear, defined as the difference
between the magnetic azimuths of the observed field and a potential
field, around the magnetic neutral lines (Hagyard et al. 1984) and
fast magnetic flux emergence (Schmieder et al. 1994) may play an
important role in their formation. These studies were mainly based on
low to moderate resolution observations.

The structure of smaller flares is much less clear since an in-depth
study requires higher resolution observations, particularly,
magnetograms. The importance of studying small flares is given by
their much higher frequency of occurrence, which scales with a power
law of index $-1.8$ (Lin et al. 1984) on the released energy. Thus,
they may provide clues on the mechanisms causing micro- and nanoflares
since the thermal events found in the corona scale with the same power
law (Wang et al. 2006), but see Pauluhn \& Solanki (2007). There are
many other reasons that the study of smaller flares is relevant for
understanding the coronal heating (see Aschwanden et al. 2007 and the
review by Krucker, 2002).

The present study is based on the fortunate coincidence, that during a
multi-wavelength observation run of a mature sunspot, a small flare
erupted above its penumbra. We present time series of high-resolution
filtergrams, together with vector magnetograms of hitherto
unprecedented spatial resolution. The photospheric magnetic field and
flow structure is analysed and related to the scenario for flaring in
the lower chromosphere.

%
%----------------------------------------------------------------
%

\section{Observations and data reduction}

\subsection{Observations}

%
%----------------------------------------------------------------
%

   \begin{figure*}
   \centering
  
   %\vspace*{2mm}

%   \parbox[b]{14cm}{
   \hspace*{12mm}\includegraphics[width=13cm]{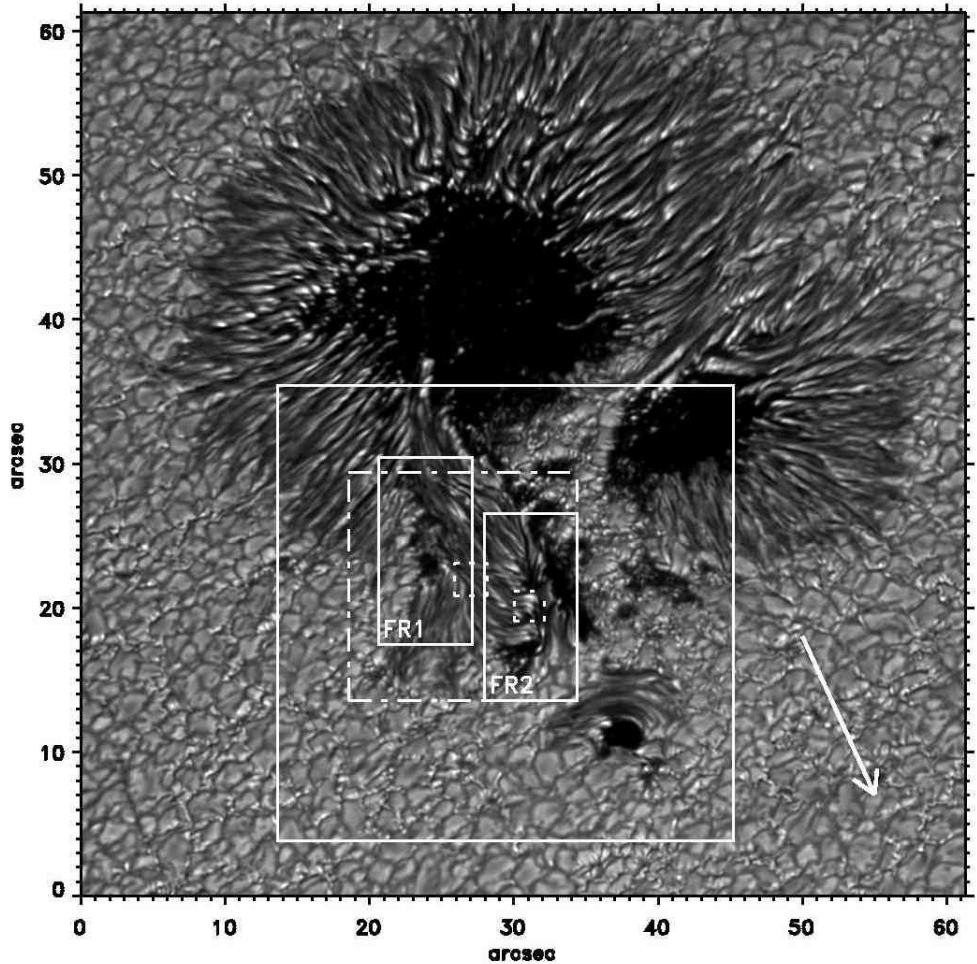}
%    }\hfill\parbox[b]{3.7cm}{
%   \includegraphics[width=3.7cm]{fig1a.eps}
%
   %\vspace*{8mm}

   \caption{Continuum image (broad band at 6302\,\AA ) of the trailing
   sunspot in NOAA 10904 on August 13, 2006, 8:44:31\,UT.  The solid
   square marks the disturbed penumbra region as displayed in
   Figs.~\ref{fig2}, \ref{fig3}, \ref{fig5}--\ref{fig7}, \ref{fig9},
   \ref{fig10}, and \ref{fig13}. The two solid rectangular boxes mark
   the two flare regions shown in Figs.~\ref{fig14}--\ref{fig16}
   (right) and Figs.~\ref{fig17}--\ref{fig19} (left). The dash-dotted
   square marks the subregions displayed in Fig.~\ref{fig12} and the
   small dotted squares show the subfields displayed in Figs.~\ref{fig8}
   (left) and~\ref{fig21} (right).  The white arrow points towards
   disk centre.}\label{fig1} 
 %  }

   \vspace*{5mm}

   \end{figure*}
%
%----------------------------------------------------------------
%

On August 13, 2006 multi-wavelength observations of the trailing
sunspot of the active region NOAA~10904 (see Figs.~\ref{fig1}
and~\ref{fig23}) were carried out at the Swedish Solar Telescope (SST)
in La Palma, Canary Islands (Scharmer et al. 2003). The centre of the
field of view was located at solar disk coordinates
$(x=-556^{\prime\prime}$, $y=-254^{\prime\prime}$), which corresponds
to a heliocentric angle of $\theta = 40.15^{\circ}$ ($\mu = 0.76$).

%
%----------------------------------------------------------------
%

   \begin{figure*}
   \centering
   \hspace*{10mm}\includegraphics[width=16cm]{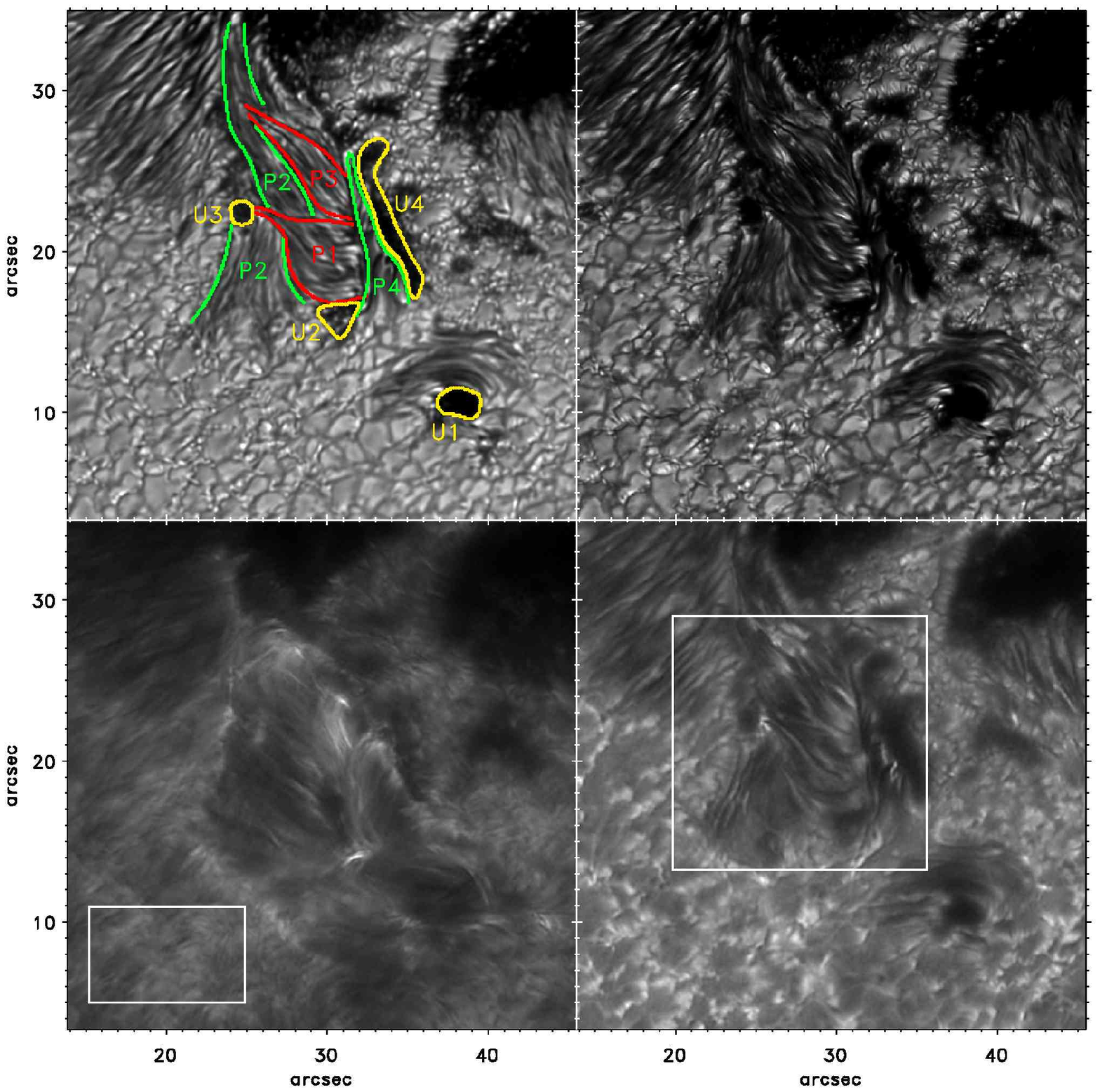} 

   %\vspace*{10mm}

   \caption{Region of interest as observed on August 13, 2006 at the
   onset of the flare. Upper left: Red continuum at 8:45:11\,UT, upper
   right: G-continuum, lower left: \ion{Ca}{ii}\,$H$ line centre,
   lower right: \ion{Ca}{ii}\,$H-0.6$\,\AA . The images from the blue
   channel were obtained at 8:44:59\,UT. In the red continuum, image
   the individual branches of the disturbed penumbra region and four
   small umbrae are outlined. The penumbral branch P2 is divided into
   two parts. The region inside of the white box of the
   \ion{Ca}{ii}\,$H$ line-centre image is used to determine the mean
   \ion{Ca}{ii}\,$H$ brightness of the ``quiet'' Sun; the white box in
   the \ion{Ca}{ii}\,$H$ line-wing image marks the subregion shown in
   Fig.~\ref{fig12}. Coordinates are given relative to the origin of
   Fig.~\ref{fig1}.}
   \label{fig2}
   \end{figure*}
%
%----------------------------------------------------------------
%

Close to the $F/47$ focus of the SST, the sunlight was divided into a
blue and a red channel by using a dichroic mirror plate. The
instrumental setup for the blue beam was arranged with four Kodak
Megaplus 1.6 CCD cameras and various interference filters. To obtain
information about the lower photosphere, one of the cameras was fed
light at the bandhead of the CH molecule at $\lambda = 4305$\,\AA$\pm
6.5$\,\AA\ (G-Band), and two cameras were fed with the light passing
an 11\,\AA\ interference filter with a central wavelength of $
4363$\,\AA\ (G-continuum). One of these cameras was set slightly out
of focus to allow subsequent image reconstruction by means of
phase-diversity wavefront sensing. (The corresponding data were not
used for the present study.) Information on the upper photosphere and
the lower chromosphere was obtained using a tiltable narrow-band
filter ($FWHM = 1.1$\,\AA) in front of the 4th camera. By tilting this
filter, the calcium \ion{Ca}{ii}\,$H$ 3968.5\,\AA\ line was recorded
at two wavelength points, once in the line centre and once in the blue
wing, approximately $0.6$\,\AA\ out of the line centre. The two
wavelength positions were alternated after recording four frames at a
time. The exposure time was set to 13\,ms for all four cameras in the
blue beam, and the achieved frame rate was between 3 and 4 frames per
second (G-band, G-continuum). Scanning the \ion{Ca}{ii}\,$H$-line
needed slightly less than 6\,s. In the blue beam, the size of one
pixel corresponds to $0\arcsec 041$.

%
%----------------------------------------------------------------
%

   \begin{figure*}
   \centering
   \hspace*{10mm}\includegraphics[width=16cm]{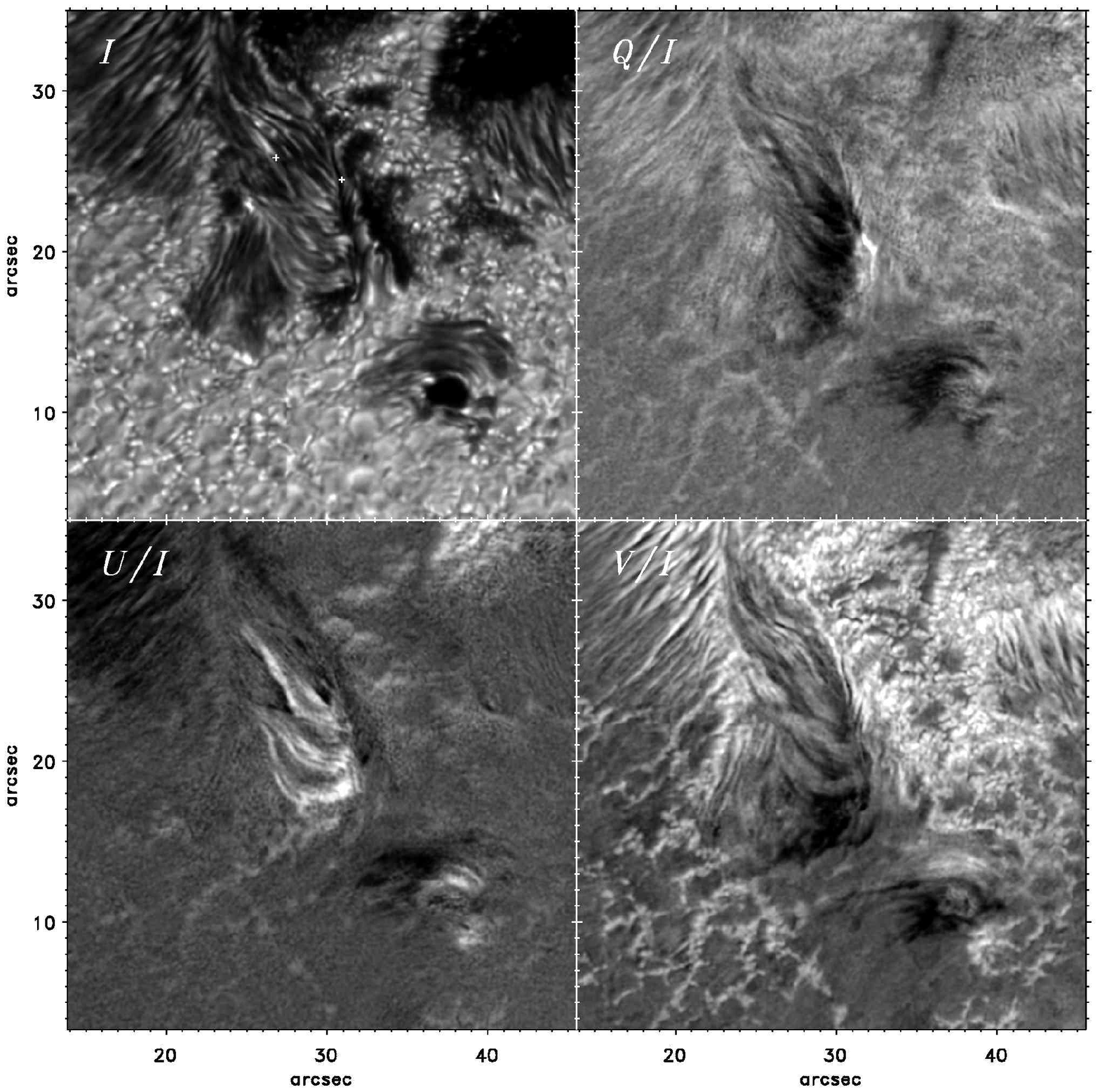} 

   %\vspace*{10mm}

   \caption{Maps of the Stokes vector of the region of interest at
   $\lambda = \lambda_0 - 75$\,m\AA . Grey levels have been scaled to
   the minimum and maximum values of each component. The `+' signs in
   the Stokes~$I$ image mark the positions of the Stokes profiles
   displayed in Fig.~\ref{fig4}.}\label{fig3}
   \end{figure*}
%
%----------------------------------------------------------------
%

The red channel was equipped with three Sarnoff CAM1M100 cameras,
which were operated at a frame rate of 36\,s$^{-1}$. The Solar Optical
Universal Polarimeter (SOUP, see Title \& Rosenberg, 1981) filter was
used to scan the iron \ion{Fe}{I} 6302.5\,\AA\ line at 6 wavelength
positions ($\lambda_0 -\lambda = [-250,-150,-75,0,75,150]$\,m\AA ).
Full Stokes polarimetry was performed by modulating the beam with two
liquid crystal variable retarder (LCVRs), which allows the full
magnetic field vector to be measured at each pixel of the field of
view. In addition, two cameras were used to obtain continuum light
(broad band) information at $\lambda = 6302$\,\AA . Again, one of
these two cameras was slightly out of focus and the corresponding data
were not used for the present study. The exposure time in the red beam
was set to 4.5\,ms by using a rotating shutter that rotated with a
speed of approximately 36\,s$^{-1}$ and opens the beam during one
sixth of each rotation. The remaining 23\,ms of each shutter rotation
were needed for detector read out and storage of the frames. The three
cameras were operated in a master-slave mode so that the exposures
were strictly simultaneous. At each wavelength position, 500 frames
were recorded at any one time. The four polarisation states were
alternated from one frame to the next. Tuning the SOUP instrument
required several seconds for each wavelength position so that a full
scan across the iron lines needed about 123\,s. The pixel size in the
red beam corresponds to $0\arcsec 065$.

The seeing conditions were excellent during a period of more than four
hours, so that the SST adaptive optics system was able to lock on the
observed region almost uninterruptedly from the start of observations
at 8:04:55\,UT. Unfortunately, the SOUP filter system was suffering
from several software problems, so that scanning of the iron line was
started only at 8:28:51\,UT and several interruptions had to be
accepted.
 
\subsection{Image reconstruction}

The G-band and the G-continuum data were bundled into packages of 60
frames. Residual influences of the Earth's atmosphere were restored by
reconstructing each of these packages with the help of speckle
interferometric techniques (Weigelt 1977; Pehlemann \& von der L\"uhe
1989; de\,Boer 1996). To consider the systematically decreasing image
quality from the lock point of the adaptive optics system towards the
borders of the frames, subfields with sizes of $128\times 128$\,pixels
were reconstructed separately with different speckle transfer
functions. The thus achieved cadence is 19\,s between subsequent
reconstructed images.
 
The slow scanning mechanism made the frame rate in the
\ion{Ca}{ii}\,$H$ channel much smaller than in the other blue
channels. To obtain statistical significance for applying speckle
reconstruction, the corresponding data were bundled into packages of
36 frames. I.e., data from nine consecutive scans were merged. By
applying this procedure, a cadence of 57\,s between subsequent
reconstructed images for each line position was achieved. In addition,
a higher cadence (19\,s) was obtained during the flare eruption by
speckle-reconstructing packages of only 12 frames (three consecutive
scans). The results of this latter procedure are only marginally worse
than those using the threefold number of frames.

%
%----------------------------------------------------------------
%

   \begin{figure}
   \centering
   \hspace*{10mm}\includegraphics[width=8cm]{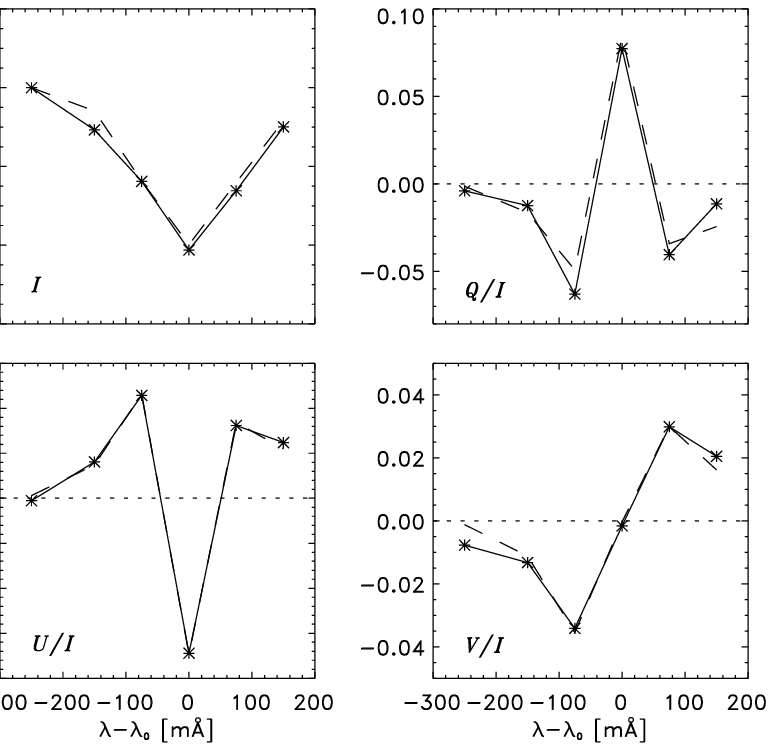} 

   \vspace*{2mm}

   \hspace*{10mm}\includegraphics[width=8cm]{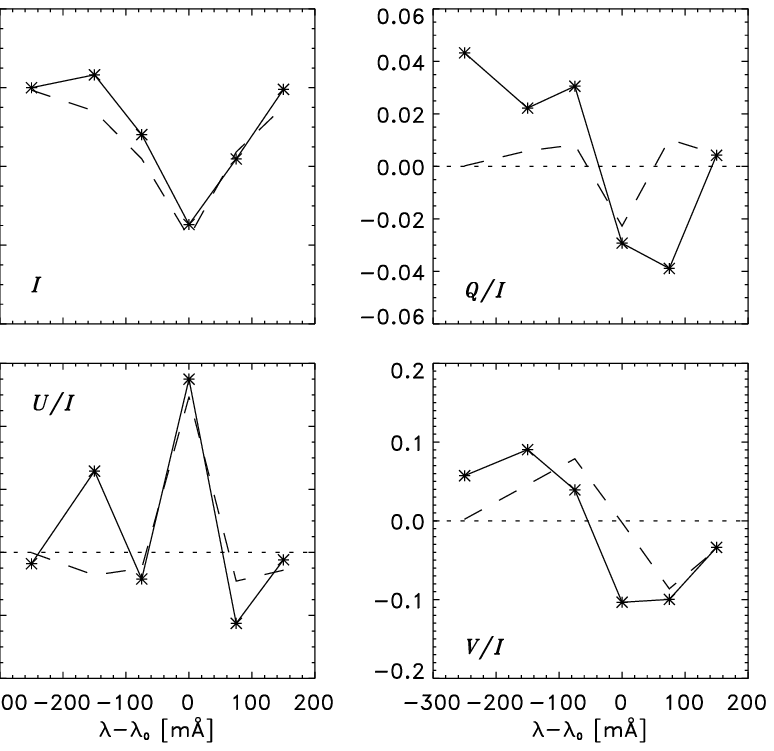} 

   \caption{Examples of regular and abnormal Stokes profiles after
   image reconstruction and demodulation as measured with the employed
   instrumental setup (solid) overplotted by the result of the
   Milne-Eddington inversion (dashed). The examples stem from a
   penumbral filament (upper four panels) and from a dark channel
   crossing the disturbed penumbra region (lower four panels; see '+'
   signs in Fig.~\ref{fig3}). Remarkable is the strong Stokes $V$
   signal at $\lambda = \lambda_0-250$\,m\AA\ in the lower example.
   This indicates an atmospheric component of strong upward mass
   flow. The one-component Milne-Eddington inversion yields a good
   approximation to the regular Stokes profiles in the upper example
   but fails to reproduce the abnormal profiles of the lower
   example. The impression of asymmetric Milne-Eddington
   profiles is caused only by the asymmetric sampling of the spectral
   line.}
   \label{fig4}
   \end{figure}
%
%----------------------------------------------------------------
%

Although, in principle, 125 images per polarisation state and line
position should have been stored in the red beam, only a much lower
number of images were fully free of interference, therefore, packages
of 50 frames were bundled in the red channels. For the camera mounted
behind the SOUP instrument, this corresponds to the 50 frames with the
highest $RMS$-contrast at each line position and polarisation
state. From the broad band channel, 50 frames obtained exactly
simultaneously with the narrow band images were selected, and speckle
reconstruction was applied. Figure~\ref{fig2} shows examples of
reconstructed broad band images (red continuum and G-continuum), as
well as \ion{Ca}{ii}\,$H$ line-core and line-wing images of the region
harbouring the flare.

Reconstruction of narrow band data (SOUP channel) was carried out by
implicitly calculating the optical transfer functions from the broad
band data (see Krieg et al. 1999). This procedure yields a 4-component
polarised light vector, ${\bm I}(x,y,t,\lambda)$, depending on the
spectral position, $\lambda$, time, $t$, and image coordinates, $x$
and $y$.

\subsection{Polarimetric demodulation}

Instrumental polarisation effects of the laboratory setup were
measured by inserting calibration optics into the beam and gradually
rotating a linear polariser. From the thus obtained data the
demodulation matrix, ${\bm M}_{\rm lab}$, was determined by using a
code developed by Selbing (2005). The instrumental polarisation of the
SST varies with the rotating field of view of the turret system. The
corresponding demodulation matrices, ${\bm M}_{\rm SST}(t)$, were
obtained by using a telescope model also provided by Selbing
(2005). Subsequently, the Stokes vectors, ${\bm S}^{\rm T}=(I,Q,U,V)$,
were computed with
\begin{equation}
{\bm S}(x,y,t,\lambda) = {\bm M}_{\rm SST}(t) \cdot 
                         {\bm M}_{\rm lab} \cdot 
                         {\bm I}(x,y,t,\lambda)\,.
\end{equation}

Examples of the resulting Stokes vectors are shown in Figs.~\ref{fig3}
and~4. The estimated $RMS$-polarimetric noise level, derived from
a comaprison of the noise of Stokes $Q$ and $U$ with Stokes $I$ in the
quiet continuum, is of the order of 1\,\%.

\section{Inversions of Stokes profiles}

To determine the magnetic field vector reliably, we inverted the
radiative transfer equation (RTE) for polarised radiation in the
Zeeman-split \ion{Fe}{i}\,6302.5\,\AA\ line. The small number of only
six wavelength points in general give us 21 significant data points
($Q$, $U$, and $V$ are usually close to zero at $\lambda_0 -\lambda =
-250$\,\AA , see however Fig.~\ref{fig4}) per spatial pixel, whereby
in the penumbra all four Stokes parameters are well above the
noise. So few observables can easily lead to unrealistic inversion
results if the number of free parameters is not kept low.

We therefore inverted the data assuming a simple
one-component-plus-straylight Milne-Eddington atmosphere using the
``HeLIx'' inversion procedure (Lagg et al. 2004). The model atmosphere
has eight free parameters. These are the three components of the
magnetic field vector ${\bm B}=(\left|{\bm B}\right|,\gamma,\chi )$ --
where $\gamma$ denotes the field inclination to the line-of-sight
(LOS) and $\chi$ is the field azimuth angle -- the LOS flow velocity,
$v_{\rm LOS}$, the Doppler broadening, the amplitudes of the
components of the propagation matrix and, finally, the geometric
filling factor, $\alpha$. In addition, a straylight atmospheric
component with a filling factor $(1-\alpha )$ was included into the
model atmosphere. In this component the magnetic field vector is
assumed to be zero, and all other parameters are coupled to the values
of the magnetic component. To keep the number of free parameters small
in the inversions, we abandoned fitting the constant term of the
source function of the Milne-Eddingtton atmospheric model. It has been
shown by Orozco Suarez et al. (2007) that this cobstraint introduces
an ambiguity between the field strengths, $\left|\bm{B}\right|$, and
the filling factor, $\alpha$, i.e. profiles produced from regions with
high field strength and low filling factor often cannot be
distinguished from those of low field strength and high filling
factor. Therefore, we henceforth use the term ``field strength''
synonymously with the flux density, $\left|\bm{B}\right|\cdot\alpha$.

%
%----------------------------------------------------------------
%

   \begin{figure*}
   \centering

%   %\vspace*{40mm}

   \hspace*{10mm}\includegraphics[width=16cm]{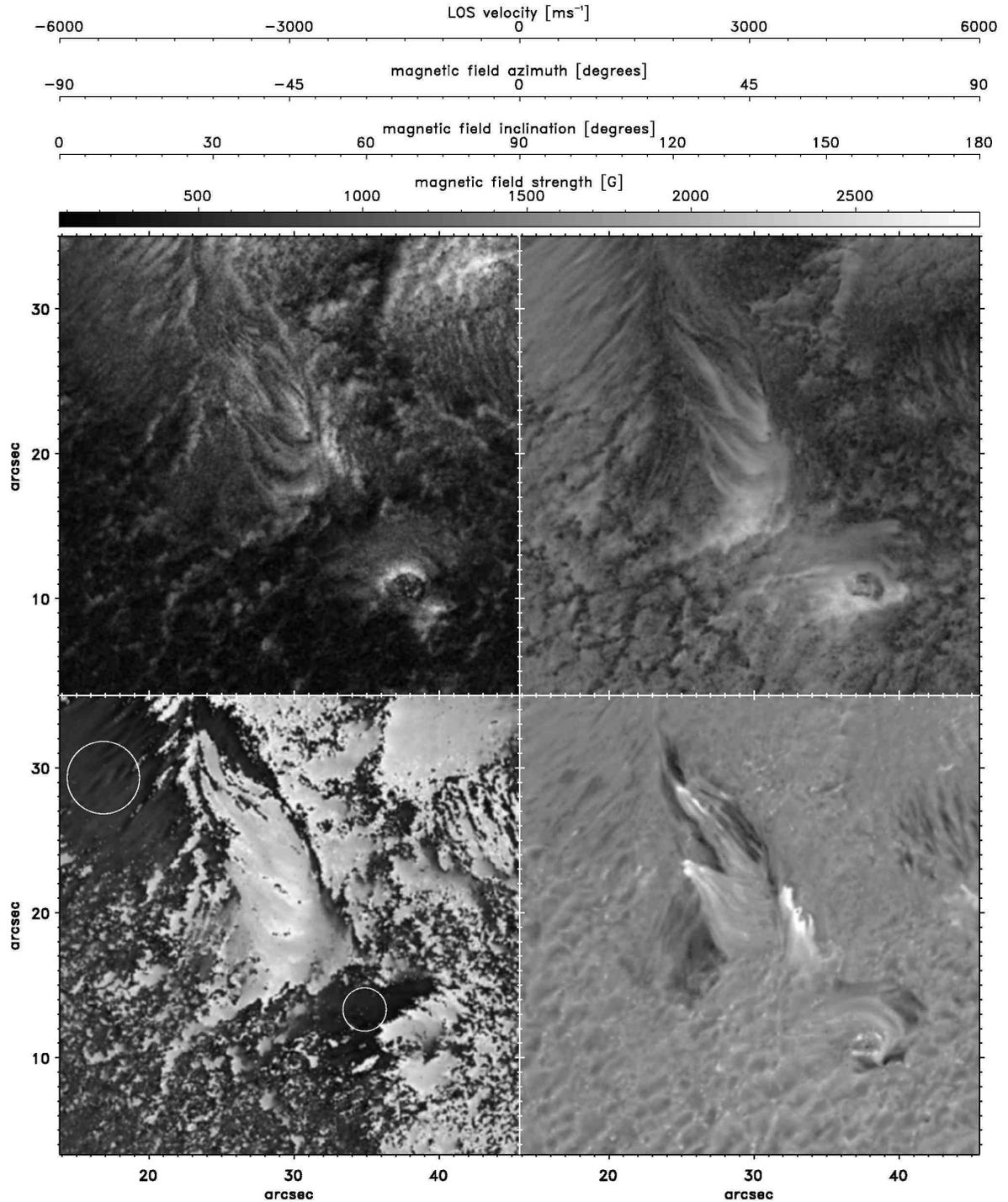} 

%   %\vspace*{10mm}

   \caption{Results of a one-component Milne-Eddington inversion of
   the Stokes profiles measured between 8:41:43\,UT and
   8:43:24\,UT. Upper left: magnetic field strength; upper right:
   field inclination relative to the LOS; lower left: field azimuth
   (perpendicular to the LOS); lower right: LOS-velocity. The regions
   marked with circles are in all likelihood suffering from an
   $180^{\circ}$ error of the field azimuth.}
   \label{fig5}
   \end{figure*}
%
%----------------------------------------------------------------
%

%
%----------------------------------------------------------------
%

   \begin{figure*}
   \centering

   %\vspace*{35mm}

   \hspace*{10mm}\includegraphics[width=16cm]{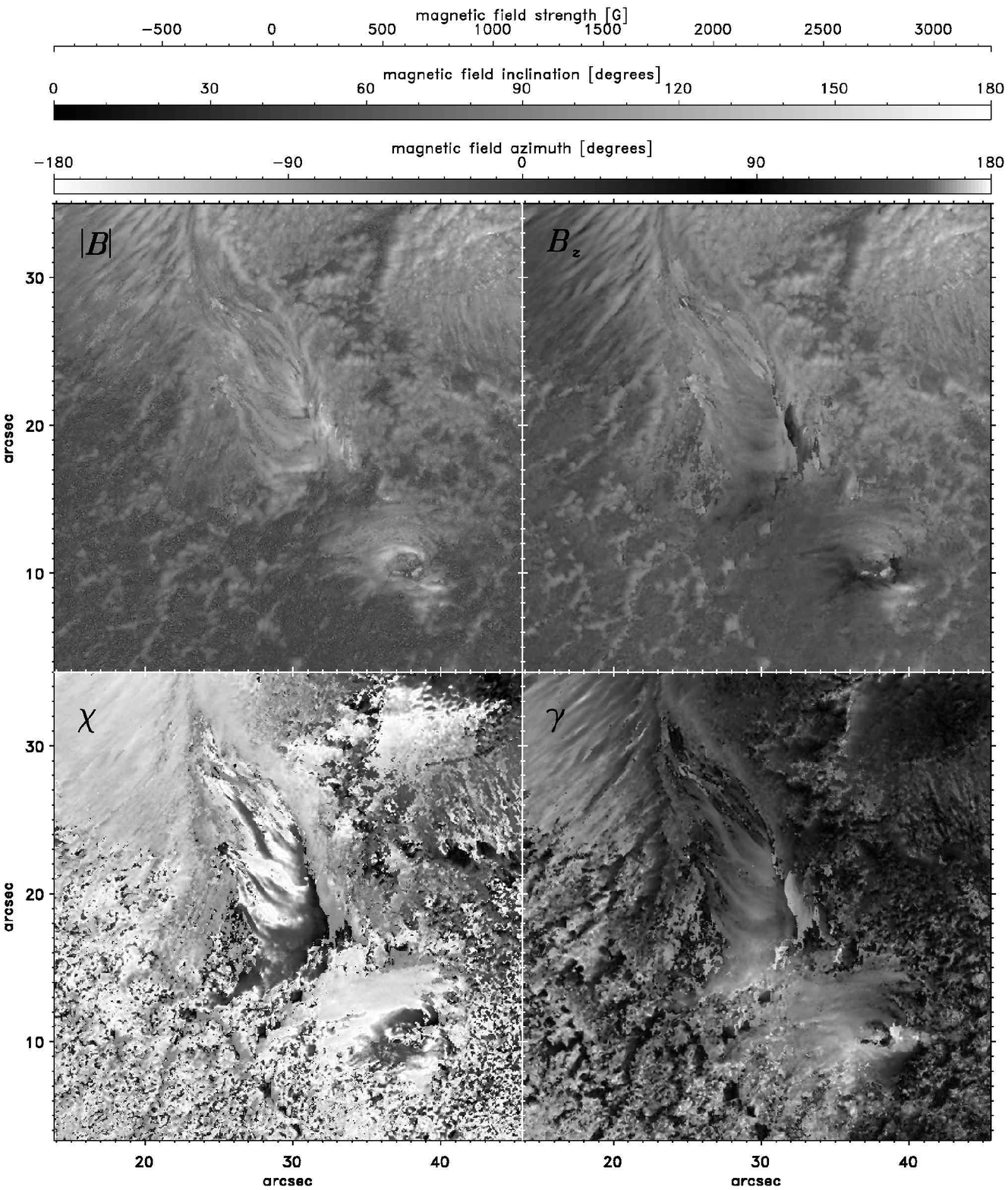} 

   %\vspace*{10mm}

   \caption{Results of a two-component Milne-Eddington inversion of
   the Stokes profiles measured between 8:41:43\,UT and 8:43:24\,UT
   after transformation into a local coordinate system where $z$ is
   perpendicular to the solar surface and $x$ , i.e. $\chi =0$, is
   directed towards the right. Upper left: $\left|{\bm
   B}\right|\alpha$, upper right: $B_z$, lower left: $\chi$, lower
   right: $\gamma$.}
   \label{fig6}
   \end{figure*}
%
%----------------------------------------------------------------
%

Selected model parameters obtained for the region of interest are
shown in Fig.~\ref{fig5}. Direct interpretation of these parameter
maps is hindered by the inevitable $180^{\circ}$-ambiguity of the
magnetic field azimuth and by the magnetic field vectors being given
in a coordinate system that is inclined to the local coordinate system
by the heliocentric angle. However, to reliably transform the field
vectors into a local coordinate system the azimuth ambiguity has to be
resolved. As can be seen in Fig.~\ref{fig5}, two large regions are
obviously afflicted with an erroneous azimuth direction. In some other
areas of the region of interest, the ``true'' azimuth direction is
less evident. Common procedures to resolve this problem (see
e.g. Metcalf et al. 2006) are often based on additional assumptions,
e.g. potentiality or minimum energy configurations, on the magnetic
field structure, which are not fulfilled in a flaring region. We,
therefore, tried to resolve the azimuth ambiguity by inverting the
azimuthal direction in those regions where it is obviously wrong and
then iteratively minimising the gradients of the field azimuth in the
region of interest (see also Zakharov et al. 2008). This procedure
works in a similar fashion to the AZAM code described by Lites et
al. (1995). In Fig.~\ref{fig6} we display the magnetic field vector
after resolution of the azimuth ambiguity and after transformation
into a local coordinate system where the $z$-direction is
perpendicular to the solar surface.

\section{Results}

Solar active region NOAA~10904 is a bipolar sunspot group comprising
two mature spots, one at each magnetic polarity. The observed trailing
spot shows mainly positive polarity (see Fig.~\ref{fig23}) and
contains tiny regions of inverse (negative) polarity in its outer
parts. Although this spot includes both polarities, it might be
classified as a part of a $\beta$-group (Potsdam classification,
cf. K\" unzel 1960) rather than as a $\delta$-spot, but see, e.g.,
Lites et al. (1995) and references therein for a discussion about the
formation of $\delta$-spots. While observing the sunspot, a flare
erupted in the disturbed penumbral region at its disc-centre side. The
flare started at about 8:47\,UT and lasted only about ten minutes. The
flare was also registered by the Solar X-ray Imager (SXI) onboard
NASA's GOES-12 satellite. The SXI data show an emission peak at
8:53:00\,UT reaching a magnitude of B7.8. In Fig.~\ref{fig2} the
disturbed penumbral region is displayed just before the start of
the flare. In the red continuum image, several penumbral branches and
small umbrae are highlighted. These features are, henceforth, called
``P1'' to ``P4'' and ``U1'' to ``U4'', respectively.

%
%----------------------------------------------------------------
%

   \begin{figure*}
   \centering

   %\vspace*{15mm}

   \hspace*{0mm}\includegraphics[width=18cm]{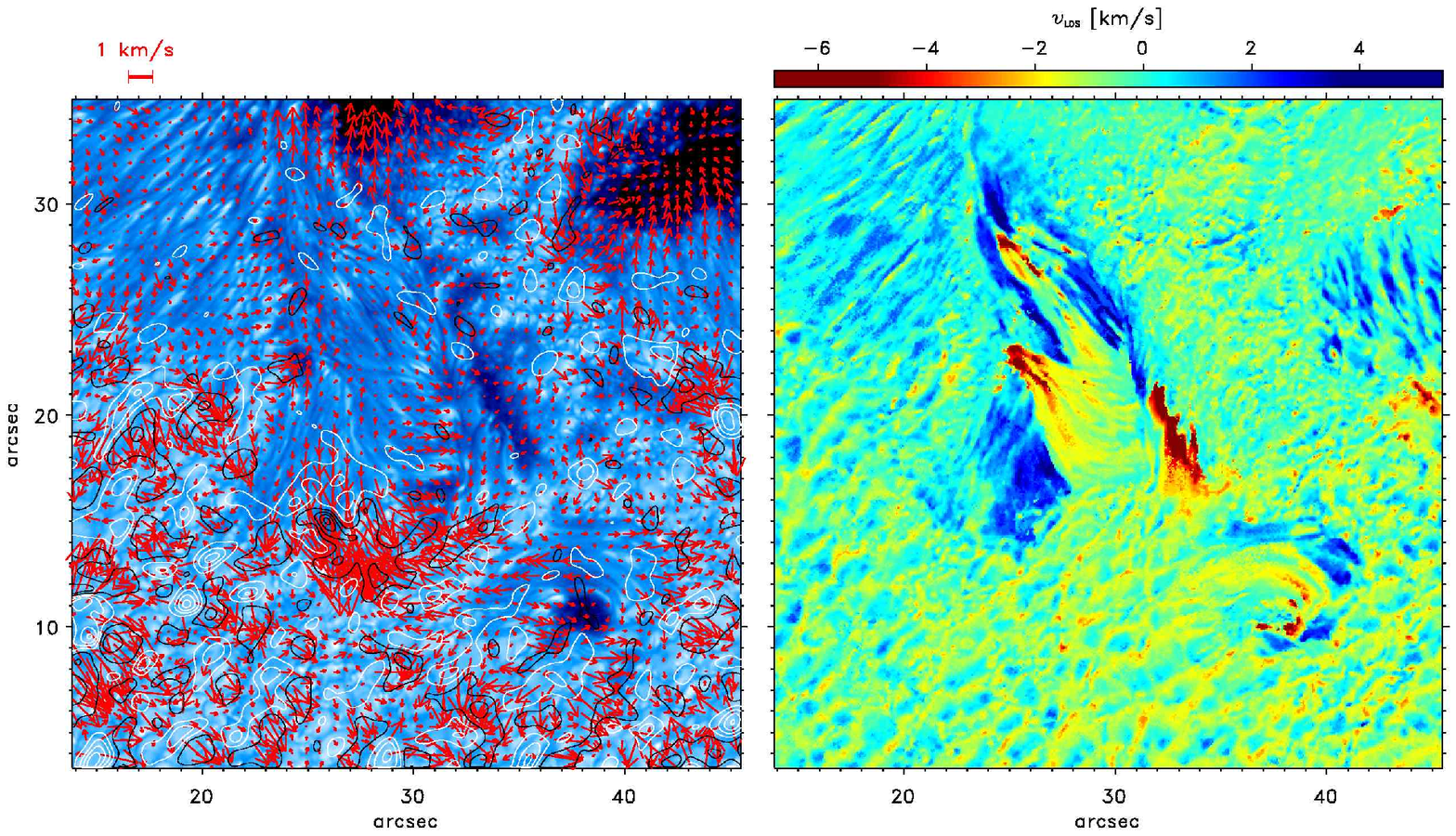} 

   %\vspace*{10mm}

   \caption{Photospheric flows in the flaring region. Left panel:
   Time-averaged horizontal flows from a series of 106 G-continuum
   images obtained between 8:28:37\,UT and 9:01:44\,UT; contour lines
   represent the divergence of the horizontal velocities (white
   contours encompass regions of positive divergence, negative
   divergence regions are located within black contours). An
   accompanying animation of the G-continuum images is available
   online at EDP.  Right panel: line-of-sight flow velocities obtained
   from a Milne-Eddington inversion of a line scan lasting from
   8:43:47\,UT to 8:45:31\,UT.}\label{fig7}
   \end{figure*}
%
%----------------------------------------------------------------
%

%
%----------------------------------------------------------------
%

   \begin{figure*}
   \centering

   %\vspace*{3mm}

   \hspace*{0mm}\includegraphics[width=18cm]{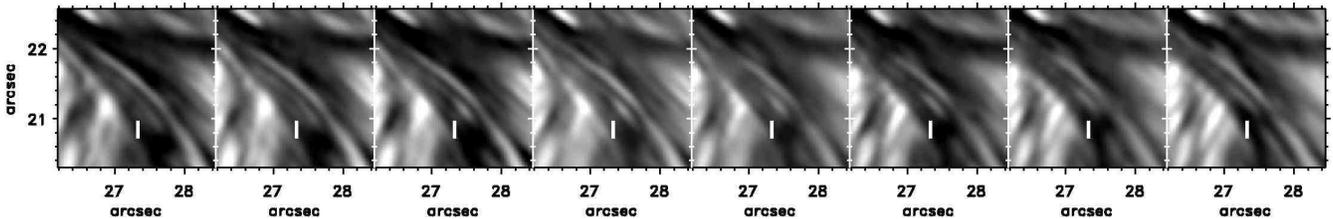} 

   %\vspace*{6mm}

   \caption{Fast motion of a fragment of a penumbral fibril located
   above the white reference marks: The figure represents a time series
   of G-continuum images between 8:38:03\,UT and
   8:40:15\,UT.}\label{fig8}
   \end{figure*}
%
%----------------------------------------------------------------
%

%
%----------------------------------------------------------------
%

   \begin{figure*}
   \centering

   %\vspace*{15mm}

   \hspace*{0mm}\includegraphics[width=18cm]{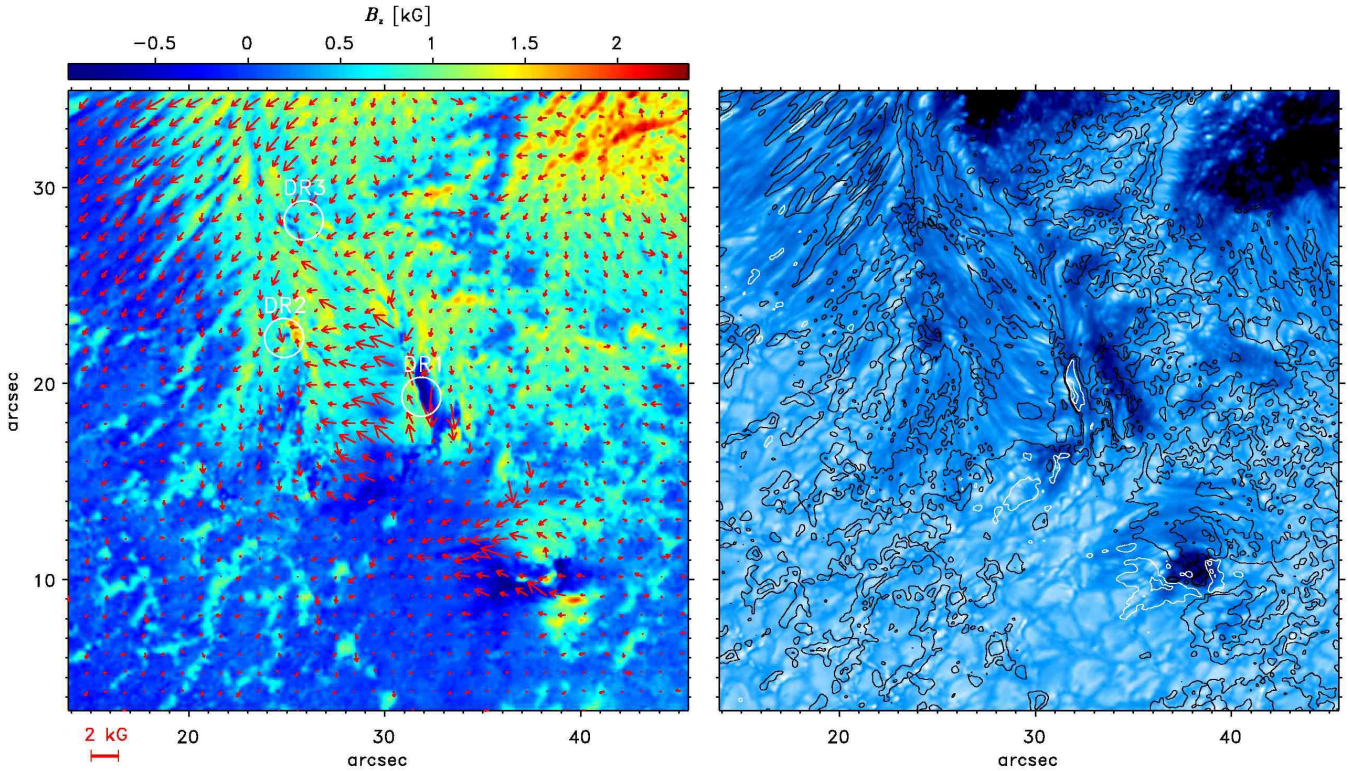} 

   %\vspace*{10mm}

   \caption{Magnetic field configuration in the flare region. Left
   panel: vertical magnetic field map ($B_z$) overplotted by vectors
   showing the horizontal field structure ($B_x$, $B_y$) obtained from
   an inversion of the line scan between 8:43:47\,UT and 8:45:31\,UT.
   Right panel: G-continuum image from 8:44:40\,UT overplotted with
   $B_z$-contour lines at $-750$, $-250$, $250$, and $750$\,G.}
   \label{fig9}
   \end{figure*}
%
%----------------------------------------------------------------
%

%
%----------------------------------------------------------------
%

   \begin{figure*}
   \centering

   %\vspace*{5mm}

   \hspace*{1mm}\includegraphics[width=18cm]{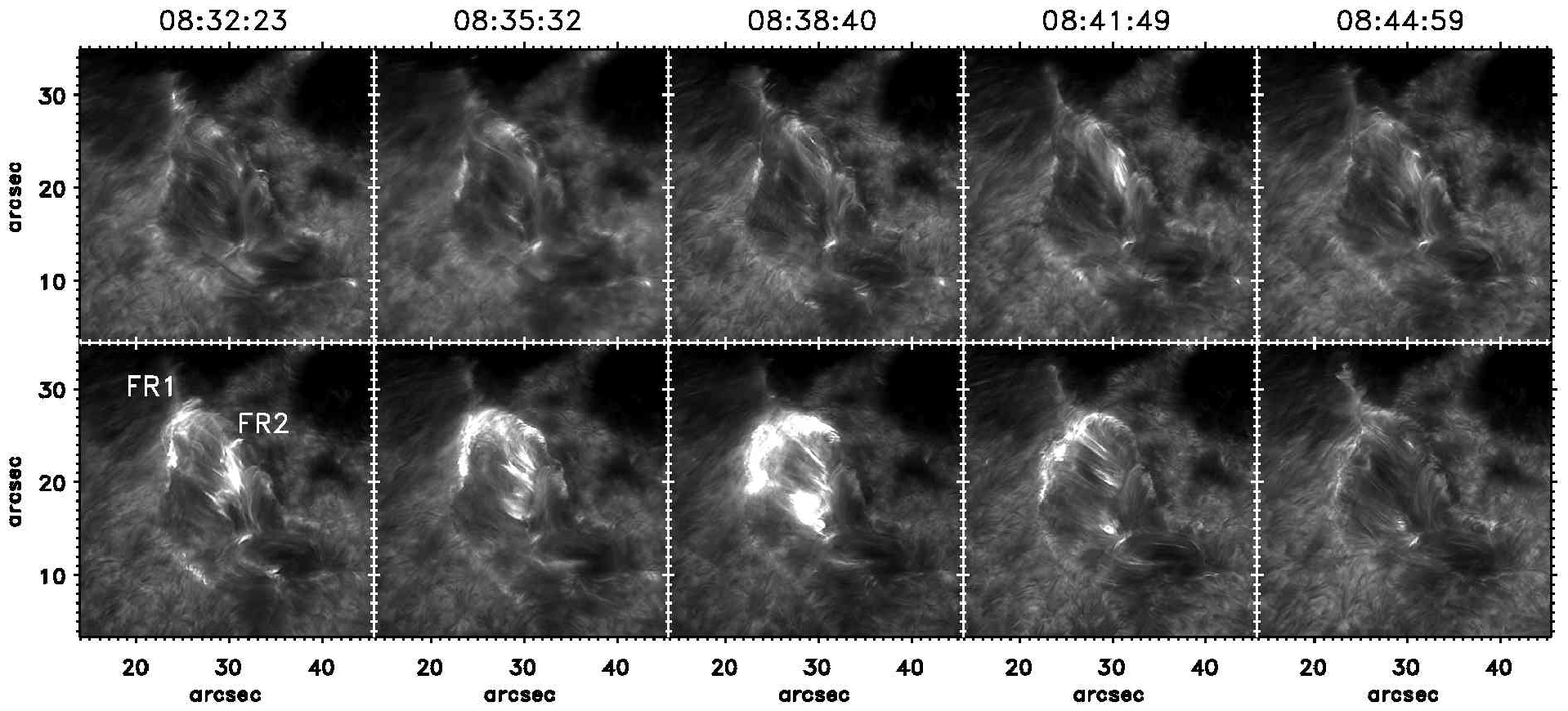} 

   %\vspace*{12mm}

   \caption{Evolution of the \ion{Ca}{ii}\,$H$ line core intensity
   during the flare eruption above the disturbed penumbra
   region. Clock time labels are given in UT. The grey code is the
   same for all images. An accompanying animation of the
   \ion{Ca}{ii}\,$H$ line core images is available online at
   EDP.}\label{fig10}
   \end{figure*}
%
%----------------------------------------------------------------
%

\subsection{Photospheric flow morphology}

Determination of the photospheric flow structure is limited by the
ability of only the LOS component being derived directly from the
Doppler shifts of spectral line profiles. The transversal flow
components can only be estimated by tracking the motions of intensity
structures, which might be misled by moving intensity patterns, such
as waves, which are not real mass flows. In the left panel of
Fig.~\ref{fig7}, the mean horizontal flow velocities in the flare
region is displayed, estimated from a 33\,min time series of
G-continuum images. The flows have been calculated using the local
correlation tracking technique (LCT, cf. November \& Simon 1988) with
a Gaussian window function of one arcsecond width.

Close to the main umbra of the sunspot the flow field shows the
well-known penetration of penumbral grains into the umbra (see
e.g. Sobotka \& S\" utterlin 2001; Bovelet \& Wiehr 2003). A
conspicuous feature in the disturbed penumbra is a fast penetration
into the adjacent quiet region at $[26\varcsec ,16\varcsec]$. There
the derived horizontal flow speeds achieve values up to
2.1\,km\,s$^{-1}$. The nearby quiet region $[26\varcsec ,15\varcsec ]$
marks a position of strong flow convergence. In the penetrating
penumbral branch (henceforth called P2, cf. Fig.~\ref{fig2}) itself
(at $[27\arcsec 5, 17\varcsec ]$), a divergence centre can be
detected. The fast horizontal velocities are not artifacts produced by
the LCT procedure since the penumbral outflow can be easily visualised
by an animation of the time series of G-continuum images, which is
available as an accompanying mpeg-movie. The concomitance of the
outward flow with continuous positive Doppler shifts (see right panel
of Fig.~\ref{fig7}) of up to $3$\,km\,s$^{-1}$ suggests fast-moving
Doppler clouds that are clearly visible in a though discontinuous
animation of the available Dopplergrams. The derived velocities of up
to 2.1\,km\,s$^{-1}$ are well within the range of the proper motion
speeds of Evershed clouds reported in Cabrera Solana et al. (2007).

In addition to the convergence centre at $[26\varcsec ,15\varcsec ]$,
other centres of horizontal flow convergence and divergence, which can
be associated with outward and inward motions, can be found all around
the disturbed penumbra. The derived flow velocities are mainly low
within the disturbed penumbra. One exception is the region around
$[31\arcsec 5,19\varcsec]$ where several bright heads of penumbral
fibrils (outlined as P1 in Fig.~\ref{fig2}) move with flow velocity of
up to 375\,m\,s$^{-1}$ towards the adjacent penumbral branch (P4),
marking another centre of flow convergence. Other exceptions are the
convergence centres at $[26\arcsec 5,28\arcsec 5]$ (close to the upper
end of branch P3) with maximum inflow speeds of about
250\,m\,s$^{-1}$, and the region left to the small umbra U3
$[23\arcsec 5,22\varcsec]$ with maximum flow speeds of
480\,m\,s$^{-1}$.

LCT can provide only temporally averaged and smoothed (by the used
window function) flows. However, inspection of a fast animated movie
of the time series reveals a wide variety of dynamic features that are
not visible in the LCT flow map; e.g., at the crossing region of
penumbral branches P1 and P2 (at $[27\arcsec 5,21\varcsec]$), the LCT
map shows a weak flow in $y$-direction. However, the corresponding
G-continuum movie shows a fast motion of the lowermost fibrils of
branch P1 towards the dark umbra U3 located at $[25\varcsec ,
22\varcsec]$; i.e., they move at an angle of more than $45^{\circ}$ to
the mean flow in that region. This motion is illustrated in
Fig.~\ref{fig8}, which figure shows a time series of $2\arcsec
25\times2\arcsec 25$ excerpts of eight consecutive G-continuum
images. Therein the motion of a fragment of a penumbral fibril can be
followed. Within the first 6 images, the structure moves about
$0\arcsec 25$, which corresponds to a velocity of
1.9\,km\,s$^{-1}$. Later on, the identification of the structure
becomes difficult. For comparison, penumbral grains move with mean
speeds between 0.5\,km\,s$^{-1}$ and 0.8\,km\,s$^{-1}$ (Sobotka \& S\"
utterlin 2001) .

As obtained from a one-component Milne-Eddington inversion of a line
scan just before the impulsive phase of the flare, the LOS velocities,
$v_{\rm LOS}$, are shown in the right panel of Fig.~\ref{fig7}. Flows
towards the observer are assigned to positive velocity values. The
entire disturbed penumbra is interspersed with exceptionally high
velocities both towards (upflows) and away from the observer
(downflows). In several regions, the downflows reach almost $v_{\rm
LOS}=-7$\,km\,s$^{-1}$, which might be higher than the local sound
speed in these regions. The detected motions in the disturbed penumbra
might be attributed to Evershed flows. Thus, the different branches of
the disturbed penumbra are characterised by Evershed motions that
point in a direction specific to each branch. Particularly interesting
is the branch P1 (located around $[27\varcsec ,22\varcsec]$) that is
associated with strong redshifts. It intersects another branch (P2)
ranging from $[24\varcsec ,30\varcsec]$ down to $[25\varcsec ,
13\varcsec]$. Therfore, P2, which is associated with strong
blueshifts, is divided into two parts (see Fig.~\ref{fig2}). Another
interesting feature is the redshifted branch P3 that intersects a
large blueshifted area around $[25\varcsec ,27\varcsec]$. Finally, one
more region associated with possibly supersonic downflows is branch
P4, located around $[33\varcsec ,19\varcsec]$. This branch corresponds
to a penumbral region consisting of short diffuse fibrils located
between an elongated umbral structure (U4) and a dark lane that
divides P4 from P1 and P3. Surprisingly, only the lower part of P4 is
redshifted, and its upper part shows blueshifts of several kilometres
per second. A strikingly twisting penumbral filament (see Ichimoto et
al. 2007, Zakharov et al. 2008) is located at the transition from red-
to blueshifts .

Supersonic flow motions in disturbed penumbral regions or in the
normal penumbra are well-documented (see Mart\'\i nez Pillet et
al. 1994 and references therein). These authors report photospheric
flow velocities (downflows) of up to more than 14\,km\,s$^{-1}$, close
to the magnetic inversion line of a $\delta$-spot. Lites et al. (2002)
found velocities of about 6 to 8\,km\,s$^{-1}$ in a broad region
between the magnetic inversion line of a $\delta$-spot and the nearby
border where upflows are separated from downflows. They hypothesised
an explanation for this result by assuming a region of interlaced
penumbral fibrils of different penumbral branches. Bellot Rubio et
al. (2007) have found evidence that the Evershed flow is supersonic at
many locations in the penumbra, confirming earlier hints (e.g. Borrero
et al. 2005).

Summarising the discussion of the features visible in the flow maps, 
it might be stated that both horizontal and LOS velocities show strong
fluctuations within short length scales. This means that the flows in
the entire disturbed penumbra seem to be strongly sheared.

\subsection{Magnetic field structure}  

The magnetic field structure of the flare region is displayed in
Fig.~\ref{fig9}. The Cartesian components of the magnetic field vector
$(B_x,B_y,B_z)$ are shown after resolving the azimuth ambiguity and
transformation into a local horizontal coordinate system. We stress
that a resolution of the azimuth ambiguity was not possible without
leaving discontinuities of the field azimuth. The applied procedure
only provides a best guess of the horizontal field direction obtained
by keeping the discontinuity regions as small as
possible. Figure~\ref{fig9} shows discontinuities at $[32\varcsec ,
19\arcsec 5]$ (DR1) and (less pronounced) at $[25\varcsec ,22\arcsec
5]$ (DR2). In the right panel of Fig.~\ref{fig9} it can be seen that
DR1 coincides with a neutral line of $B_z$, separating a small region
of negative (downward) vertical field from the surrounding regions of
positive (upward) $B_z$. The second discontinuity region (DR2) is
located at the right border of the small umbra U3.

Penumbral branch P1 (see Fig.~\ref{fig2}) connects these two magnetic
field discontinuities and carries fast negative Evershed motion from
DR1 to DR2 (see Fig.~\ref{fig7}). The magnetic field in P1 is mainly
horizontal and directed towards DR2. This is almost perpendicular to
the radial direction of the sunspot. A close inspection of the Doppler
map and the G-continuum images (see Fig.~\ref{fig7}) indicates that
this branch crosses penumbral branch P2 at the position of DR2. The
underlying branch P2 carries a fast positive Evershed flow that is
directed more or less radially outwards. Also the field vectors point
nearly radially outwards. Thus, this branch might be considered as the
``normal'' spot penumbra that is intersected by P1, and P3 is another
branch of penumbral filaments carrying negative Evershed flows
(located between $[26\varcsec ,28\varcsec]$ and the flow convergence
region at $[31\varcsec ,19\varcsec]$). This might be a sub-branch of
P1 that also intersects the global structure of the sunspot penumbra.
Again the magnetic fields in this branch are in opposite directions to
the global fields of the sunspot. A convergence centre of the
horizontal flows is located at the upper end of this branch (i.e. at
$[26\varcsec , 28\varcsec]$). Also the horizontal magnetic field is
somewhat noisy (therefore, we call this region DR3, henceforth),
although a clear discontinuity cannot be detected. This noisy
behaviour come from the fact that a Milne-Eddington inversion assumes
a height-independent magnetic field, which is most likely not valid in
a region of crossing penumbral branches.
 
The description of the global magnetic field structure in the
disturbed penumbra region can be summarised as the fields being
largely co-aligned with the Evershed flows. This leads to
discontinuities in the horizontal fields that are co-spatial with the
convergence points of the Evershed flows and also of convergence
centres of the horizontal flows (proper motions of penumbral
structures). In addition, several islands of oppositely (downward)
directed vertical fields are present within and around the disturbed
penumbral region. This configuration is largely comparable to the
magnetic field configuration of the flare model developed in Falconer
et al. (2000).

\subsection{\ion{Ca}{ii}\,$H$ observations}

The evolution of the flare with time as seen in the \ion{Ca}{ii}\,$H$
line centre images obtained with the SST is shown in
Fig.~\ref{fig10}. An animation is available as accompanying
mpeg-movie. The pre- and post-flaring region above the disturbed
penumbra is easily discernible (see also Fig.~\ref{fig2}). It is
darker than the adjacent rather quiet region, visible in the lower
left corner of the images displayed in Fig.~\ref{fig10} and also
darker than the disturbed granular region, which appears between the
disturbed penumbra and the two umbrae of the spot. The border between
the pre-flaring region and its surroundings shows enhanced brightness
in the line core of \ion{Ca}{ii}\,$H$. This bright border might be
interpreted as the intersection of a separatrix dividing the magnetic
fields inside the disturbed penumbra from its surroundings with the
chromospheric level where the calcium line is formed (see Falconer et
al. 2000 and below).

%
%----------------------------------------------------------------
%

   \begin{figure}

   \centering

   \hspace*{-2.5mm}\includegraphics[width=7.5cm]{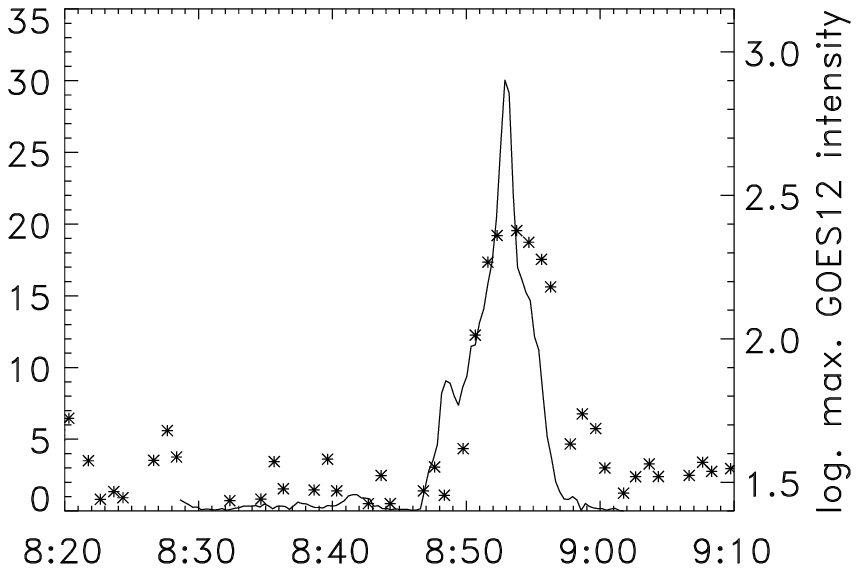} 

   \vspace*{3mm}

   \caption{Temporal variation of the flaring area, which is defined by
   those regions that are brighter than twice the mean quiet
   Sun \ion{Ca}{ii}\,$H$ line core brightness (solid). The asterisks
   show the logarithmic maximum SXI/GOES12 intensity of the flaring
   region (arbitrary units).}\label{fig11}
   \end{figure}
%
%----------------------------------------------------------------
%

Inside the region enclosed by the separatrix the \ion{Ca}{ii}\,$H$
line core images show a jumble of long fibrils that are mostly
connected to different parts of the bright border or to small-scale
bright structures inside this region. These structures might be
closely related to the ``straws'' described by Rutten (2007) and the
\ion{Ca}{ii}\,$K$ fibrils recently discovered by Pietarila et
al. (2009). This appearance is significantly different from the normal
penumbra at the limb-side of the sunspot. There the fibrils are
radially oriented, i.e. mainly parallel to the photospheric penumbral
fibrils. The normal \ion{Ca}{ii}\,$H$ line core penumbra is also
rougher-textured and -- on average -- darker than the disturbed flare
region (see Fig.~\ref{fig10online} for a \ion{Ca}{ii}\,$H$ line
core image of the entire field of view of the observations.).

As in a mature two-ribbon flare, the eruption starts simultaneously at
two stripes, which are separated by approximately 7 to 8 arcseconds
(see lower left panel of Fig.~\ref{fig10}). One of the two ribbons
(FR1) appears first just at the separatrix and extends later on to an
opposite part of the separatrix, marking the inner penumbral boundary
(sometimes called peripatopause) between the disturbed penumbra and
the larger sunspot umbra. Two branches of fibrils originally cross in
the region of the second flare ribbon (FR2). One of these branches
flashes up and expands until it merges with FR1 (see panel from
8:48:09 to 8:52:52\,UT). During the rapid expansion of the upper part
of FR2 (see panel from 8:51:18\,UT), the flashing fibrils form an
arcade of bright streaks at the upper part of the flaring region. This
arcade is located between the upper ends of the two flare ribbons. Its
upper footpoints are co-spatial with the separatrix region. Later on,
the centre of the flare brightness moves toward the lower ends of the
flare ribbons (see panel from 8:52:52\,UT) and the maximum brightness
seems to be emitted by long thin fibrils crossing and enclosing both
flare ribbons, as well as from their footpoints at FR1. The width of
these fibrils is only about 3 to 4 pixels, i.e. about $0\arcsec 15$
although their length reaches values up to 10\,arcsec. These values
are in good agreement with the dimensions of recently discovered
\ion{Ca}{ii}\,$K$ fibrils seen in plage regions (Pietarila et
al. 2009). After 8:52\,UT, the flare brightness slowly decreases until
the region reaches a similar state as during its pre-flare phase at
approximately 9:00\,UT.

Unfortunately, the total intensity flux of the flare could not be
measured since the detector was saturated during the impulsive phase
of the flare. Instead of that, we estimated the evolution of the total
flaring area (see Fig.~\ref{fig11}). This area is defined as regions
achieving intensities more than double that of the average quiet Sun
\ion{Ca}{ii}\,$H$ line core intensity estimated from a region to the
lower left of the flaring region (see Fig.~\ref{fig2}). The flaring
area reaches its maximum extent of 30.94\,Mm$^2$ at 8:52:52\,UT. Its
rising phase lasts slightly longer than its decline and, as already
mentioned, shows a few bumps. In general, the shape of this curve is
similar to that of the GOES measurements although only one smooth peak
is visible in the GOES-12 X-ray measurements. GOES-8 X-ray data show a
bump in the declining phase of the flare.

%
%----------------------------------------------------------------
%

   \begin{figure*}
   \centering

   %\vspace*{5mm}

   \hspace*{1mm}\includegraphics[width=18cm]{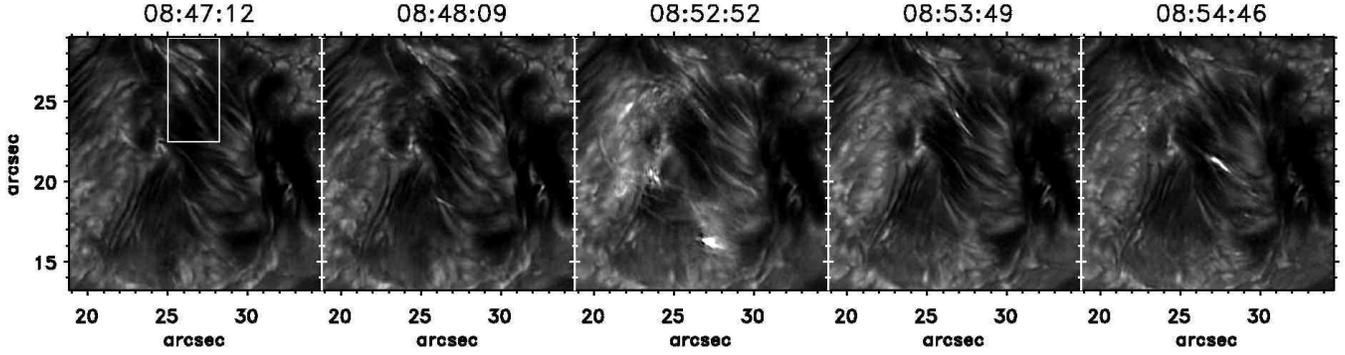} 

   %\vspace*{7mm}

   \caption{Evolution of the \ion{Ca}{ii}\,$H$ wing intensity in a
   subregion (see Fig.~\ref{fig2}). The white box marks the location
   of the detail shown in Fig.~\ref{fig20}. An accompanying animation
   of the \ion{Ca}{ii}\,$H$ line wing images is available online at
   EDP.}
   \label{fig12}
   \end{figure*}
%
%----------------------------------------------------------------
%

The effects of the flare eruption on the blue wing of
\ion{Ca}{ii}\,$H$ are shown in Fig.~\ref{fig12}. An animation is
available as an accompanying mpeg-movie. The first indication of the flare
appears at 8:47:12\,UT as an isolated bright filamentary structure (in
Fig.~\ref{fig12} at position $[22\arcsec 5,22\arcsec 5]$). At
8:48:09\,UT another bright structure appears at $[26\arcsec
5,18\arcsec 5]$, which seems to be an extension of the one seen
before. Finally, as the flare reaches its maximum brightness in the
line centre, a definite loop structure becomes visible. Its footpoints
are rooted in rather close proximity to each other. One of the
footpoints is associated with a bright flash at 8:52:52\,UT, whereas
the other one shows no brightness enhancement until 8:53:49\,UT. Later
on, another bright structure appears at 8:54:46\,UT right between the
two footpoints at $[27\arcsec 5,21\varcsec]$.

%
%----------------------------------------------------------------
%

   \begin{figure}
   \centering

   %\vspace*{2mm}

   \hspace*{0mm}\includegraphics[width=8cm]{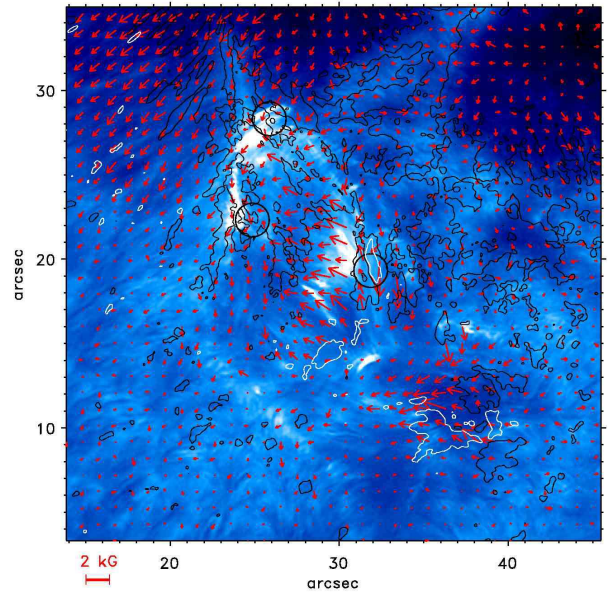} 

   %\vspace*{10mm}

   \caption{Horizontal magnetic field (arrows) and vertical magnetic
   field (contours at $-250$ and $750$\,G) overplotted to the
   \ion{Ca}{ii}\,$H$ line-core intensity obtained at 8:47:31\,UT at
   the onset flare eruption. Black circles mark regions of magnetic
   azimuth {\bf discontinuity} (cf.~Fig.~\ref{fig9}).}\label{fig13}
   \end{figure}
%
%----------------------------------------------------------------
%

\subsection{Relation of magnetic fields and the flare ribbons}

To demonstrate that the determined magnetic field structure is
realistic, the horizontal magnetic field is overplotted in
Fig.~\ref{fig13} on a \ion{Ca}{ii}\,$H$ line core image at the onset
of the flare eruption. The region of magnetic azimuth discontinuity
DR1 at $[32\varcsec , 19\arcsec 5]$ is located to the right of the
onset region of FR2. The onset of FR1 takes place above the small
umbra U3, leftward of DR2. Figure~\ref{fig13} shows that region DR3
also participates in the flare since it is located slightly to the
right of the upper end of FR1. The displacement between the
discontinuity regions and the flare ribbons might be (partially)
explained by a parallax caused by the inclined line of sight. The
displacement of the discontinuities from the local flare intensity
maxima is approximately one arcsecond in all three regions. This
implies a (reasonable) difference of 1100\,km between the height where
the photospheric magnetic field information is obtained and the height
of the brightest \ion{Ca}{ii}\,$H$ line core flashes.

\subsection{The onset of the flare}  

\begin{figure*}
   \centering

   %\vspace*{5mm}

   \hspace*{2mm}\includegraphics[width=17cm]{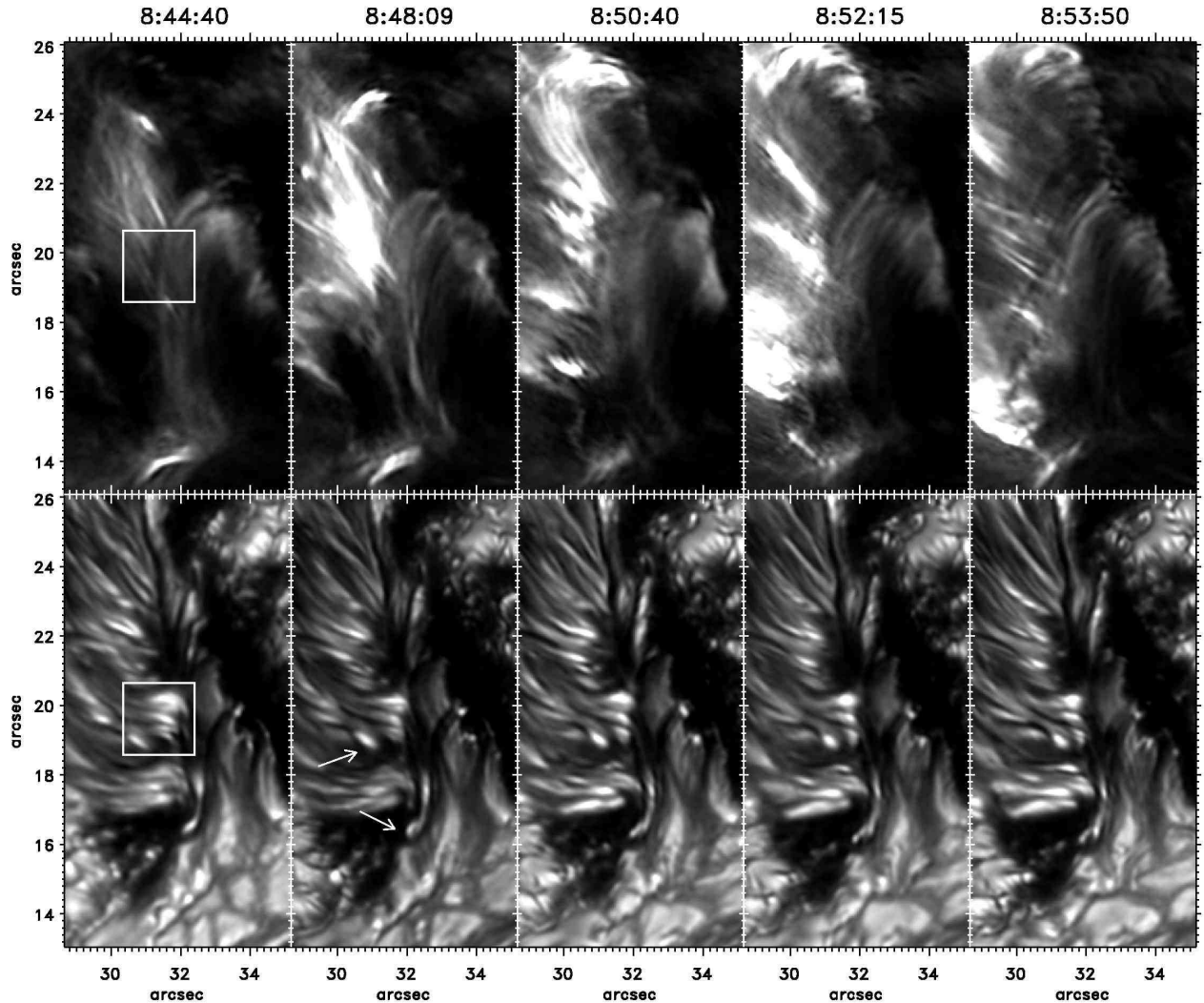} 

   %\vspace*{7mm}

   \caption{The onset of the flare in FR2 as seen in the
   \ion{Ca}{ii}\,$H$ line-core (upper panels) and in the blue
   continuum (lower panels). The white boxes mark the subregion
   displayed in Fig.~\ref{fig21}. Time labels are given in UT.}
   \label{fig14}
   \end{figure*}

The first appearance of the flare takes place in a region of two
intersecting branches of \ion{Ca}{ii}\,$H$ line-core fibrils. In
Fig.~\ref{fig14} the evolution of this region (FR2) as seen in the
\ion{Ca}{ii}\,$H$ line core and in the blue continuum is displayed.
After its first appearance, the fine structure of FR2, as displayed in
Fig.~\ref{fig14}, shows a series of bright loop-like structures (at
8:50:40\,UT), which connect the position of their first appearance with
a previously dark region. These loops expand rapidly and move towards
the second flare ribbon. After that, the region of maximum
\ion{Ca}{ii}\,$H$ brightness (cf. Fig.~\ref{fig10}) spreads towards
the lower left corner of the cutouts shown in Fig.~\ref{fig14}.

The chromospheric structure in this region cannot only be
characterised by two crossing branches of fibrils: In the blue wing of
the \ion{Ca}{ii}\,$H$ line (see Fig.~\ref{fig12}, panel for
8:52:52\,UT) the onset region of the flare is connected with the lower
end of the flare ribbon (i.e. with the small umbra (U2) in the lower
left corner of the details shown in Fig.~\ref{fig14}) via an extended
loop. Since this loop is not clearly visible in the \ion{Ca}{ii}\,$H$
line-centre images, it must be either low-lying and/or become apparent
due to strong Doppler shifts. Only moderate dynamics occur in the blue
continuum. The most conspicuous phenomena are two fast moving and
evolving penumbral filaments (Fig.~\ref{fig14}).

%
%----------------------------------------------------------------
%

%
%----------------------------------------------------------------
%

\begin{figure}
   \centering

   %\vspace*{5mm}

   \hspace*{2mm}\includegraphics[width=7.5cm]{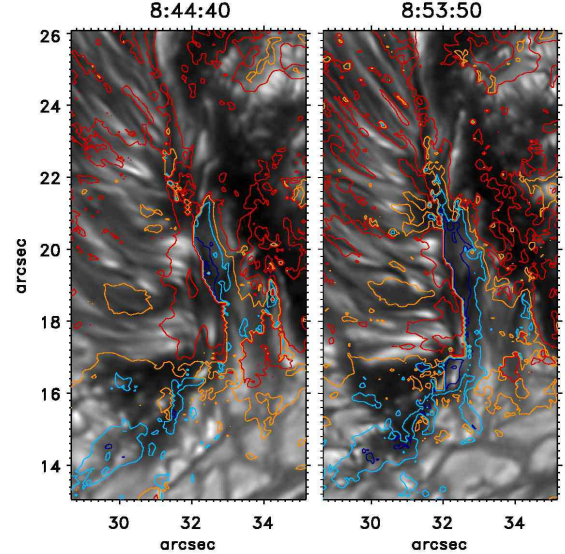} 

   %\vspace*{8mm}

   \caption{Vertical magnetic field strength in the region of FR2
   before and after the onset of the flare. Contour levels are at
   -750\,G (blue), -250\,G (green), 250\,G (yellow), and 750\,G
   (red). Background images are G-continuum data obtained at the times
   labelled above the panels.}\label{fig15}
   \end{figure}

\begin{figure}
   \centering

   %\vspace*{5mm}

   \hspace*{2mm}\includegraphics[width=7.5cm]{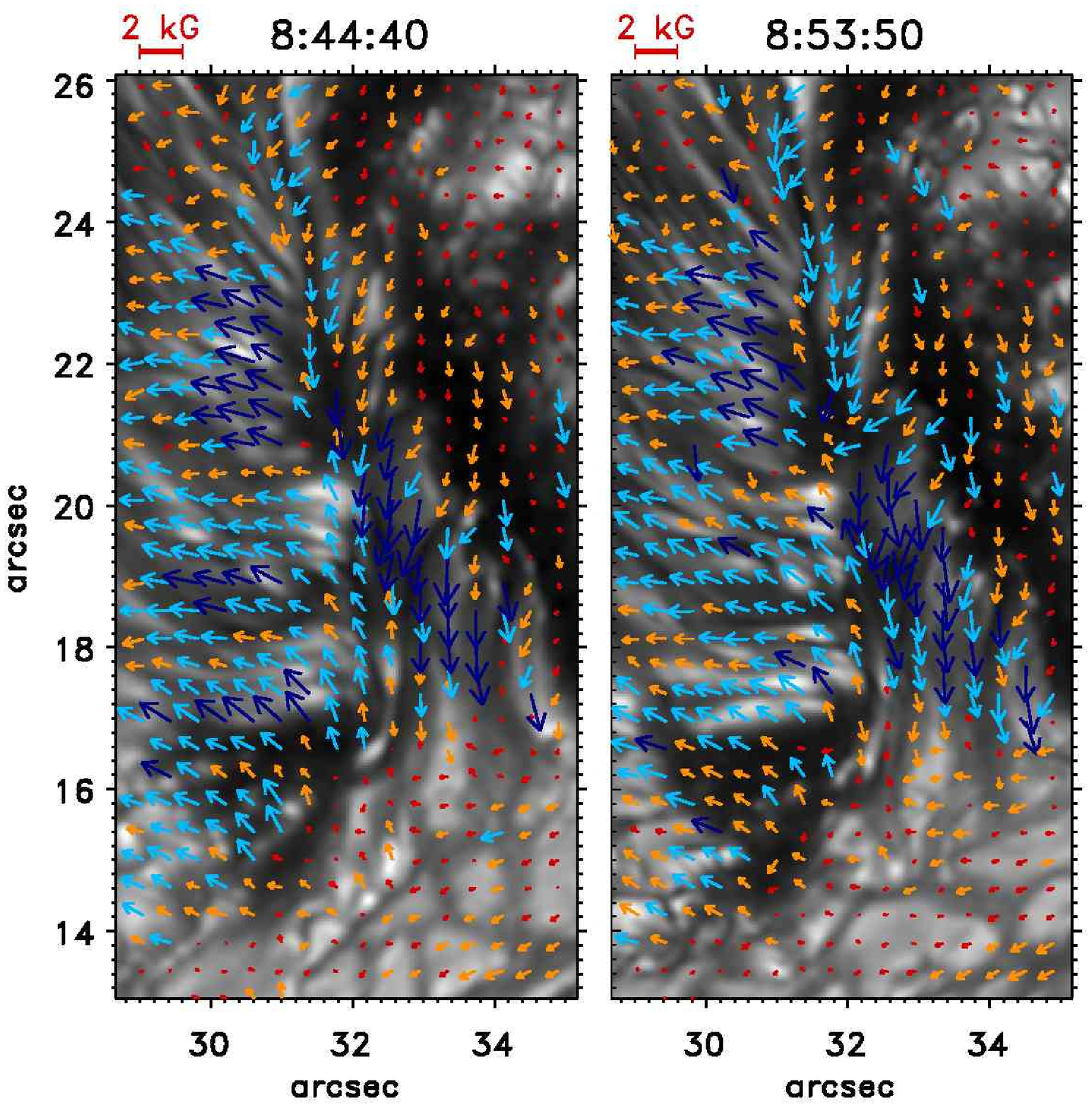} 

   %\vspace*{8mm}

   \caption{Direction and strength of the horizontal magnetic field
   strength in the region of FR2 before and after the onset of the
   flare. Arrow colours denote field strengths $B_{\rm h}<500$\,G (red),
   $500$\,G$\enspace\leq B_{\rm h}<1000$\,G (yellow),
   $1000$\,G$\enspace\leq\!B_{\rm h}<1500$\,G (green), and $B_{\rm
   h}\geq 1500$\,G (blue). Background images are G-continuum data
   obtained at the times labelled above the panels.}\label{fig16}
   \end{figure}

The corresponding magnetic field configuration is displayed in
Figs.~\ref{fig15} and~\ref{fig16}. Because of technical problems with
the SOUP, magnetic field maps are available only shortly before (at
8:44:40\,UT) and shortly after (at 8:53:50\,UT) the flare onset. The
most conspicuous feature in this region is a sharp reversal of the
sign of $B_z$ in the centre of the cutouts shown in
Fig.~\ref{fig15}. This situation is not resolved by the flare
eruption. Surprisingly, the island of negative $B_z$ grows during the
flare eruption. This growth takes place mainly in the lower part of
the displayed details where a penumbral filament (see lower white
arrow in Fig.~\ref{fig14}) moves into the small umbra U2. Throughout
the entire observed period the magnetic neutral line separates
penumbral branch P1, which carries fast Evershed motion from DR1 to
DR2, from the island of negative $B_z$ which appears as a diffuse
penumbral region (P4) in the blue continuum images. According to the
moving flux tube model (cf. Schlichenmaier et al. 1998, but see,
however, Rempel et al. 2009) for penumbral filaments, the bright heads
of the fibrils, which are located just leftward of the neutral line,
appear where vertical flux tubes bend over and become
horizontal. Thus, the field lines are mainly vertical in this region
and at least some become more inclined, i.e. $B_z$ decreases and the
horizontal field component, $B_{\rm h}=(B_x^2+B_y^2)^{1/2}$,
increases, in about one arcsecond distance from the neutral line.
Also the Doppler maps (see Fig.~\ref{fig7}) show blueshifts in the
region close to the neutral line, although the biggest fraction of this
penumbral branch is characterised by strong redshifts.

%
%----------------------------------------------------------------
%

%
%----------------------------------------------------------------
%

%
%----------------------------------------------------------------
%

%
%----------------------------------------------------------------
%

A similar situation is present just rightward of the neutral line. In
the magnetic island the $z$-component is strongest in close proximity
to the neutral line and $B_{\rm h}$ achieves values up to 2.5\,kG at a
distance of $0.5$ to 1\,arcsec from the neutral line. In the dark
umbral region (U4) rightward of the magnetic island, $B_z$ is again
strong (however positive) and $B_{\rm h}$ drops to values around
500\,G. Therefore, it might be concluded that the diffuse penumbral
region (P4) contains loop-like magnetic fields which emerge in the
dark elongated umbra (U4) that is located in the upper right part of
the cutouts shown in Figs.~\ref{fig14}--\ref{fig16} and submerge in
the island of negative $B_z$ close to the bright heads of the
penumbral filaments of P1. Thus, the magnetic neutral line represents
a region of strong vertical (and also horizontal) magnetic shear.  We
stress that the entire region of FR2, including the island of negative
$B_z$, is located inside the separatrix appearing in the
\ion{Ca}{ii}\,$H$ images. In fact, the separatrix is co-spatial with
U4.

\begin{figure*}
   \centering

   %\vspace*{5mm}

   \hspace*{2mm}\includegraphics[width=17cm]{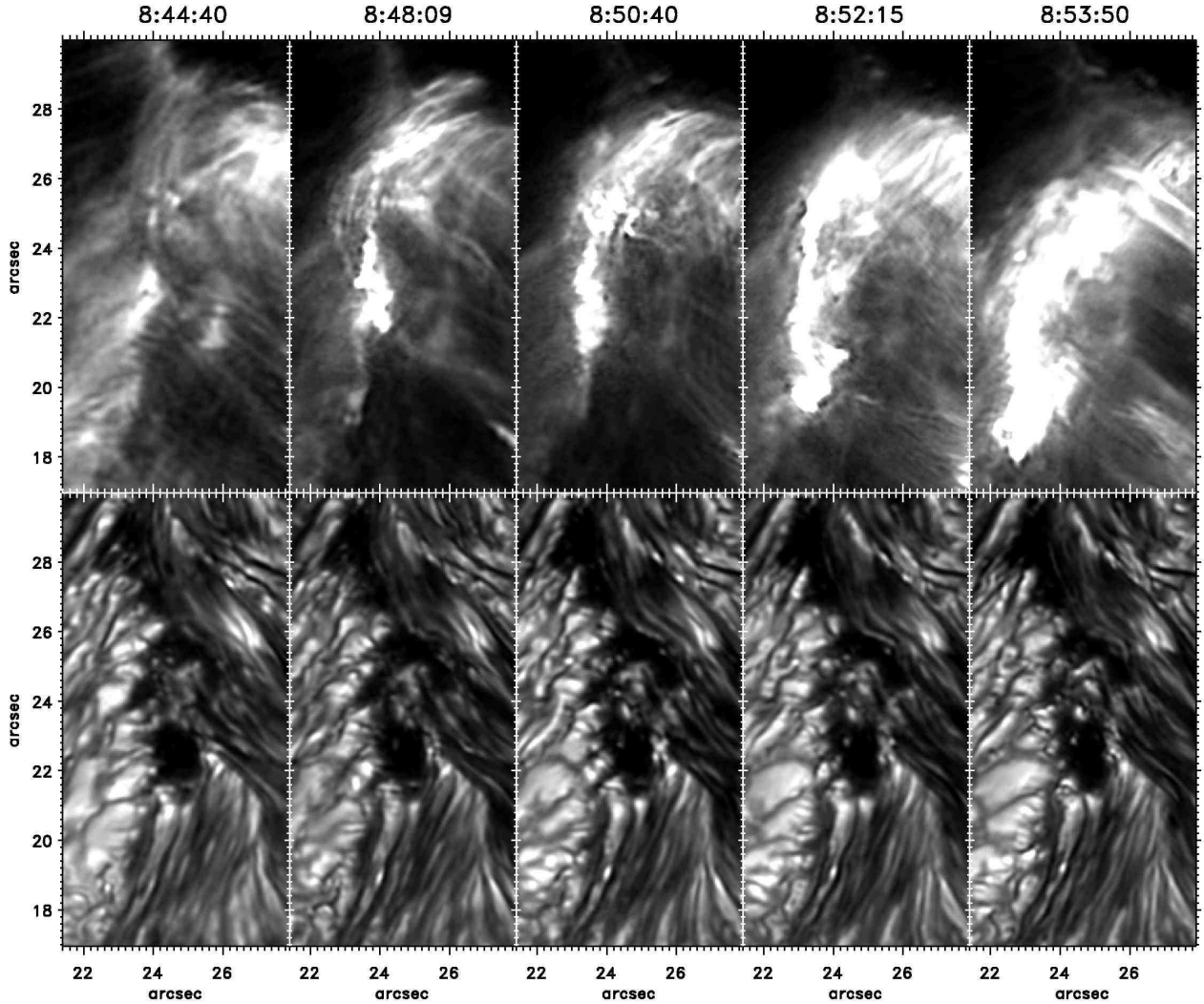} 

   %\vspace*{7mm}

   \caption{The onset of flare ribbon FR1 as seen in the
   \ion{Ca}{ii}\,$H$ line-core (upper panels) and in the blue
   continuum (lower panels). Time labels are given in
   UT.}\label{fig17}
   \end{figure*}

\begin{figure}
   \centering

   %\vspace*{5mm}

   \hspace*{2mm}\includegraphics[width=7.5cm]{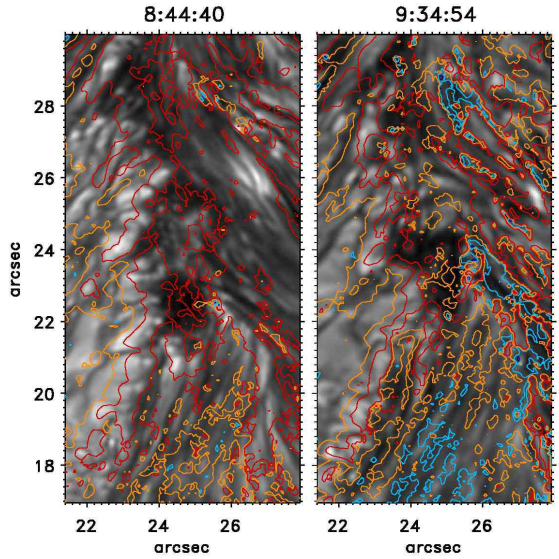} 

   %\vspace*{8mm}

   \caption{Vertical magnetic field strength around FR1 at the onset
   and 40\,min after the flare. Contour levels are at -750\,G (blue),
   -250\,G (green), 250\,G (yellow), and 750\,G (red). Background
   images are G-continuum data obtained at the times labelled above
   the panels.}\label{fig18}
   \end{figure}

\begin{figure}
   \centering

   %\vspace*{5mm}

   \hspace*{2mm}\includegraphics[width=7.5cm]{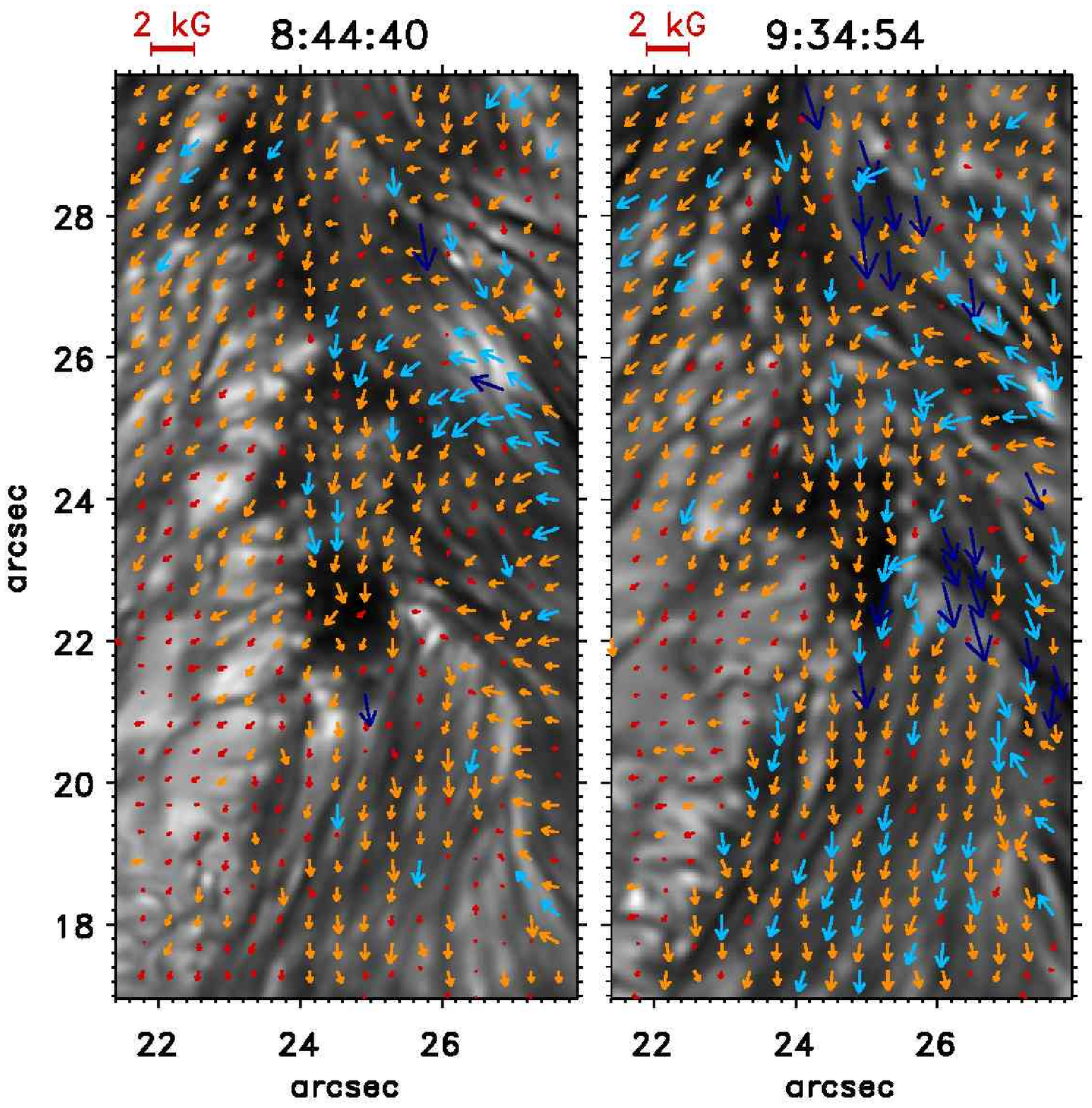} 

   %\vspace*{8mm}

   \caption{Direction and strength of the horizontal magnetic field
   strength around FR1 at the onset and 40\,min after the flare.
   Arrow colours denote field strengths $B_{\rm h}<500$\,G (red),
   $500$\,G$\enspace\leq B_{\rm h}<1000$\,G (yellow),
   $1000$\,G$\enspace\leq\!B_{\rm h}<1500$\,G (green), and $B_{\rm
   h}\geq 1500$\,G (blue). Background images are G-continuum data
   obtained at the times labelled above the panels.}\label{fig19}
   \end{figure}

The onset of the other flare ribbon (FR1) is displayed in
Fig.~\ref{fig17}. Figures~\ref{fig18} and~\ref{fig19} show the
corresponding magnetic field configuration. The flare onset takes
place above the small roundish umbra U3 (visible in the middle of
lower row frames in Fig.~\ref{fig17}) and spreads out in both
directions along the ribbon. The \ion{Ca}{ii}\,$H$ line-core images
show some loop-like fibrils which connect U3 with the region around
DR3. Again the photospheric structure shows only moderate dynamics.
The magnetic field structure at the time of the flare onset
(8:44:40\,UT) also does not show many conspicuous features. Worth
mentioning are only two tiny regions of negative $B_z$ at the right
border of the small umbra and close to DR3 (at $[25\arcsec
5,28\varcsec ]$). The right panels of Figs.~\ref{fig18}
and~\ref{fig19} show the field configuration approximately 40\,min
after the flare eruption. It can be clearly seen that the field
structure has changed dramatically. Mainly the two regions of negative
$B_z$ have become much larger and also the penumbral branch below the
small umbra (lower part of P2) mainly comprises negative
$B_z$. Simultaneously with the growth of the regions of negative
$B_z$, in the same regions the horizontal field strength strongly
increases up to values of more than 2\,kG.

%
%----------------------------------------------------------------
%

%
%----------------------------------------------------------------
%

%
%----------------------------------------------------------------
%

%
%----------------------------------------------------------------
%

%
%----------------------------------------------------------------
%

%
%----------------------------------------------------------------
%

The scenario developed from analysing the Doppler maps and the general
field topology in the region of interest can be substantiated and
extended now. In this region several penumbral branches seem to
intersect. The longest one (P2) reaches from the upper border of the
details displayed in Fig.~\ref{fig17} to their lower border. It seems
to submerge below P1 slightly above the small roundish umbra U3 and
re-emerges below this umbra. This submergence of P2 below P1 might be
also assumed from animating a time series of the blue continuum
images. The proper motions therein in both parts of branch P2 are seen
to be directed outward, whereas in the branch P1 the penumbral
substructures seem to move inward, i.e. towards DR2, on one side and
outward (towards DR1) on its other side.  Figure~\ref{fig8} shows a
typical structure moving towards DR2. The magnetic field lines in the
region below U3 (lower part of P2) must have a loop-like structure
since $B_{\rm h}$ is continuously directed downward and $B_z$ is
positive close to the U3 and turns to negative values at a distance of
a few arcseconds from U3.

During the observation sequence, the magnetic structure of this region
displays some significant changes. The explanation for these changes
that we prefer is that the regions of negative $B_z$ are emerging from
below, which is mainly reflected by a growth of the regions of
negative $B_z$, and that this field emergence triggers the flare. In
spite of these changes, the intersection of penumbral branches seems
to persist during the entire observation period.

\subsection{Dynamic features}  

\begin{figure}
   \centering

   %\vspace*{5mm}

   \hspace*{2mm}\includegraphics[width=7.5cm]{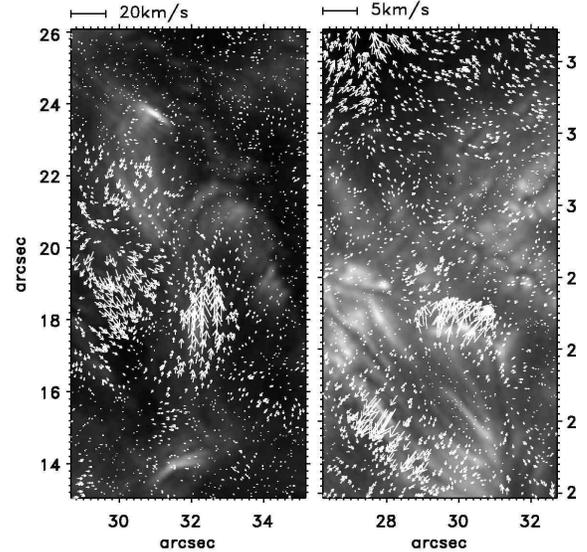} 

   %\vspace*{8mm}

   \caption{Left panel: Horizontal flow field as obtained from the
   \ion{Ca}{ii}\,$H$ line-core images between 8:45:56\,UT and
   8:48:47\,UT. The cutout displays the same region as shown in
   Figs.~\ref{fig14}--\ref{fig16}. Right panel: Horizontal flow field
   as obtained from the \ion{Ca}{ii}\,$H$ line-wing images between
   8:50:40\,UT and 8:56:41\,UT in the region marked by a white box in
   Fig.~\ref{fig12}.}\label{fig20}
   \end{figure}

A wide variety of dynamic features is visible on extremely small
(spatial and temporal) scales in our data. In Fig.~\ref{fig20} (left
panel) the region already shown in Fig.~\ref{fig14} is overplotted
with a horizontal flow field obtained from the \ion{Ca}{ii}\,$H$
images between 8:45:56\,UT and 8:48:47\,UT. Before the flare appears,
the region shows two interlaced branches of \ion{Ca}{ii}\,$H$ fibrils
(see upper left panel in Fig.~\ref{fig14}). The flow field was derived
using LCT with a Gaussian window function of $0\arcsec 9$
width. Within this time interval, a bright cloud is moving along the
branch of \ion{Ca}{ii}\,$H$ fibrils that lies above the island of
negative $B_z$ shown in Fig.~\ref{fig15}. This cloud achieves maximum
speeds of 25.5\,km\,s$^{-1}$ just as it crosses the island of negative
$B_z$ (at $[32\varcsec ,18\varcsec]$ in Fig~\ref{fig20}). This
velocity lies at the upper limit of chromospheric flows typically
observed in H$\alpha$ (see e.g. Al et al. 2004, De Pontieu et
al. 2007) or in \ion{Ca}{ii}\,8662\,\AA\ (see Langangen et al. 2008),
although van~Noort \& Rouppe van der Voort (2006) found blobs moving
with speeds up to 240\,km\,s$^{-1}$. Line-of-sight flows exceeding
this value are also regularly seen in \ion{He}{i}\,10830\,\AA\ spectra
(cf. e.g. Aznar Cuadrado et al. 2005, Lagg et al. 2007, Sasso et
al. 2009). The flare eruption takes place (first enhancement of
\ion{Ca}{ii}\,$H$ brightness in this region) approximately one minute
after the cloud has passed the crossing of the two branches of fibrils
and approximately 40\,s after the cloud reaches the upper end of the
branch.

%
%----------------------------------------------------------------
%

%
%----------------------------------------------------------------
%

\begin{figure*}
   \centering

   %\vspace*{5mm}

   \hspace*{2mm}\includegraphics[width=18cm]{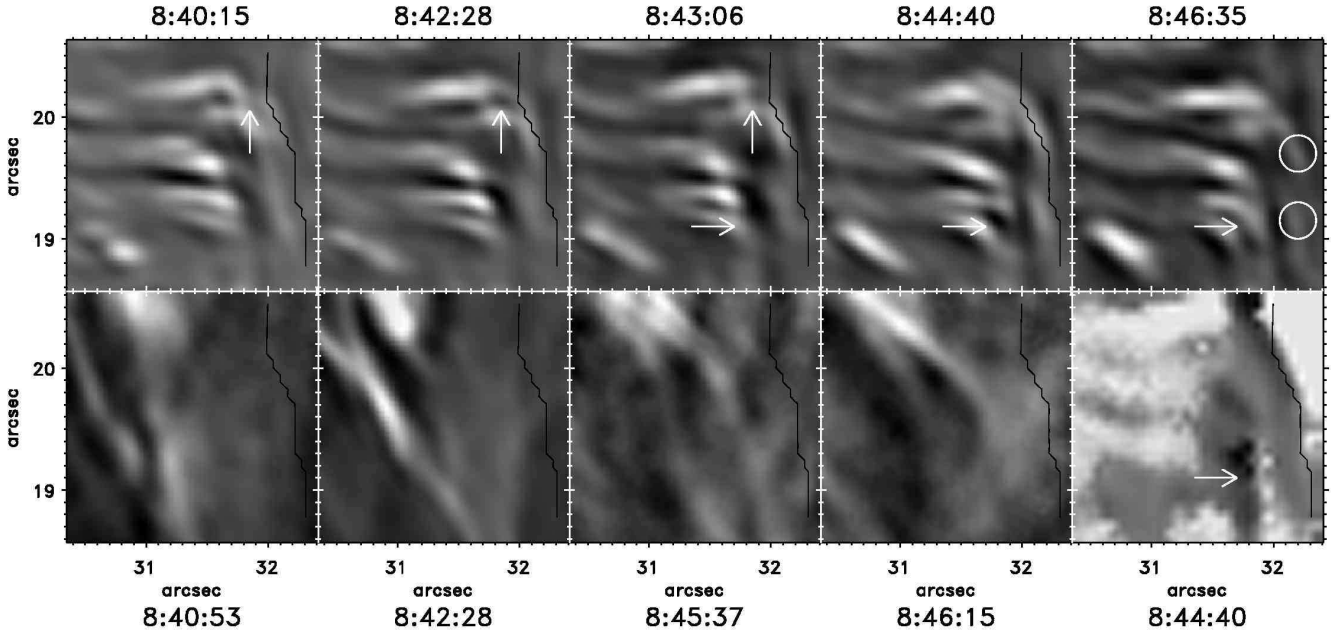} 

   %\vspace*{13mm}

   \caption{Evolution of penumbral branch P1 at the onset of the flare
   (FR2). Upper panels: contrast-enhanced G-continuum images.  Dark
   structures are crossing the dark-cored penumbral filaments (arrows
   define reference positions). Bright remnants of a decaying
   penumbral filament are marked by circles. Lower panels:
   contrast-enhanced \ion{Ca}{ii}\,$H$ line-core images and and the
   corresponding Doppler map (lower right panel) scaled to values
   between $v_{\rm LOS}=-1.5$\,km\,s$^{-1}$ (white) and $v_{\rm
   LOS}=1$\,km\,s$^{-1}$ (black). The overplotted line shows the
   magnetic neutral line ($B_z=0$) at 8:44:40\,UT. The location of
   this subregion is marked with white boxes in Figs.~\ref{fig1}
   and~\ref{fig14}. Time labels are given in UT.}
   \label{fig21}
   \end{figure*}

During the same period (i.e. approximately between 8:40\,UT and
8:47\,UT) in the photosphere below, penumbral filaments with typical
dark cores (see Scharmer et al. 2002) are located (see
Fig.~\ref{fig21}). In the uppermost filament visible in the details
shown in Fig.~\ref{fig21}, a dark roundish structure moves from the
dark core of the filament towards the magnetic neutral line (see
panels from 8:40:15\,UT to 8:43:06\,UT in Fig.~\ref{fig21}). Its size
is only about $0\arcsec15$. Later on (see panels from 8:43:06\,UT to
8:46:35\,UT in Fig.~\ref{fig21}), superposed to the lowermost of the
three filaments, another tiny but elongated dark structure
crosses. This dark structure is about $0\arcsec 15$ wide and not
longer than $0\arcsec 3$. In an animation of this subregion, these
features appear as conspicuous dark structures, leaving the filaments
unmodified. Also the corresponding Doppler maps show no anomalies. The
regions close to the centres of the bright heads of the penumbral
filaments exhibit strong blueshifts and the dark lanes between the
filaments indicate strong redshifts that can be attributed to Evershed
flows. These features are also visible in the \ion{Ca}{ii}\,$H$
line-wing images. Therefore, they might be ascribed to density
enhancements appearing in the temperature minimum or even in the lower
chromosphere. several fibrils are visible also in the
\ion{Ca}{ii}\,$H$ line-core images (see Fig.~\ref{fig14} and the lower
panels of Fig.~\ref{fig21}). During the appearance of the dark
photospheric structures, these fibrils are already increasing their
intensity (onset of the flare). In the contrast-enhanced
\ion{Ca}{ii}\,$H$ image from 8:46:15\,UT in Fig.~\ref{fig21}, the
fast-moving cloud described in the previous paragraph can also be
seen. It is located in the lower right corner of this panel above two
conspicuous bright fibrils and just crosses the magnetic neutral
line.

The uppermost of the three penumbral filaments displayed in
Fig.~\ref{fig21} shows another dynamic feature that appears as a
bright extension of this filament (Fig.~\ref{fig21}, panel from
8:46:35\,UT). This bright extension is (considering that the spatial
resolution of the magnetogram is not better than $0\arcsec 2 -
0\arcsec 3$) located exactly below the magnetic neutral line that
separates the island of negative $B_z$ from the penumbral filaments
where $B_z>0$. This structure is a remnant of a slowly decaying bright
penumbral filament located along the magnetic neutral line
approximately 25\,min prior to the flare eruption. It disappears
completely during the onset of the flare. The less conspicuous
structure marked in Fig.~\ref{fig21} is another remnant of the same
penumbral filament.

The two previously discussed features show clearly that the
photosphere below the flare undergoes a significant restructuring
shortly before the flare erupts, although only on extremely small
scales.

Another striking feature that can be seen in Fig.~\ref{fig21} is the
strongly different orientation of photospheric penumbral fibrils and
chromospheric \ion{Ca}{ii}\,$H$ fibrils. The two types of features are
misaligned by an angle of more than $45^{\circ}$. Such a misalignment
of photospheric and chromospheric structures can be seen in several
parts of this region. The region around $[26\varcsec ,26\varcsec ]$ in
the second panel (8:48:09\,UT) of Fig.~\ref{fig17} shows another
example. There the misalignment is even of the order of
$75^{\circ}$. The photospheric field is roughly aligned with the
penumbral fibrils (compare left panel of Fig.~\ref{fig9} with the
upper left panel of Fig.~\ref{fig3}). If we now assume that the
\ion{Ca}{ii}\,$H$ fibrils mark chromospheric field lines, then it
would appear that the field structure in the observed region not only
shows strong horizontal shear but also a strong vertical rotation;
i.e., the flare occurs in a region of strong helicity. The flare does
not fully remove the misalignment of chromospheric and photospheric
structures. As can be seen in the upper right panels of
Figs.~\ref{fig14} and~\ref{fig17}, intersecting branches of
\ion{Ca}{ii}\,$H$ fibrils remain in the analysed region and the
underlying penumbral fibrils remain misaligned with at least one
branch of the overlying \ion{Ca}{ii}\,$H$ fibrils. Different
orientations in the photosphere and chromosphere may also have to do
with different inclinations of photospheric and chromospheric
structures with respect to the solar surface, since the sunspot was
significantly far removed from disk centre ($\theta=40.15^\circ$) at
the time of observation.

%
%----------------------------------------------------------------
%

%
%----------------------------------------------------------------
%

Another dynamical feature, visible during the main phase of the flare,
is shown in the right panel of Fig.~\ref{fig20}. This figure shows the
\ion{Ca}{ii}\,$H$ line-wing intensity in the region between the two
flare ribbons. Overplotted is the horizontal flow field obtained from
the \ion{Ca}{ii}\,$H$ line-wing images between 8:50:40\,UT and
8:56:41\,UT. In the centre of the cutout (at $[30\arcsec 5,27\varcsec
]$; see also Fig.~\ref{fig12}, panel from 8:52:52\,UT), a bright front
is moving towards the top-right. The velocities (using LCT with a
$0\arcsec 9$ window function) achieve values up to
4.8\,km\,s$^{-1}$. It is co-spatial with the outer border of the
rapidly expanding flare loop visible in the \ion{Ca}{ii}\,$H$
line-centre images (see Fig.~\ref{fig14}) and is, thus, not exactly
co-spatial with the brightest parts of the line-core images. The width
of this front is only approximately $0\arcsec 25$. If this bright
front were only caused by the contribution of the line-centre
intensities to the line-wing images (since the filter width is larger
than the wavelength difference between the two channels), it should be
co-spatial with the brightest parts of the flaring region. Although
the detector was saturated in some regions, this is definitely not the
case. We, therefore, assume a strong blueshift of the
\ion{Ca}{ii}\,$H$ line in the front and also in the fibrils flashing
in the line-wing images. Using the wavelength difference of 0.6\,\AA\
between the two channels results in a speed of 45\,km\,s$^{-1}$.

A second region of fast horizontal motions is located in the lower
left corner of the right panel of Fig.~\ref{fig20}. These motions are
related to fibrils appearing bright in the \ion{Ca}{ii}\,$H$
line-wing. Since a distinct moving wavefront is not visible, it
remains unclear if the bright fibrils are really moving or whether we
only see various fibrils flashing at different times that mislead the
LCT algorithm.

\section{Discussions and conclusions}

\begin{figure*}
   \centering

   \vspace*{-30mm}
   \hspace*{-18mm}\includegraphics[width=21.5cm]{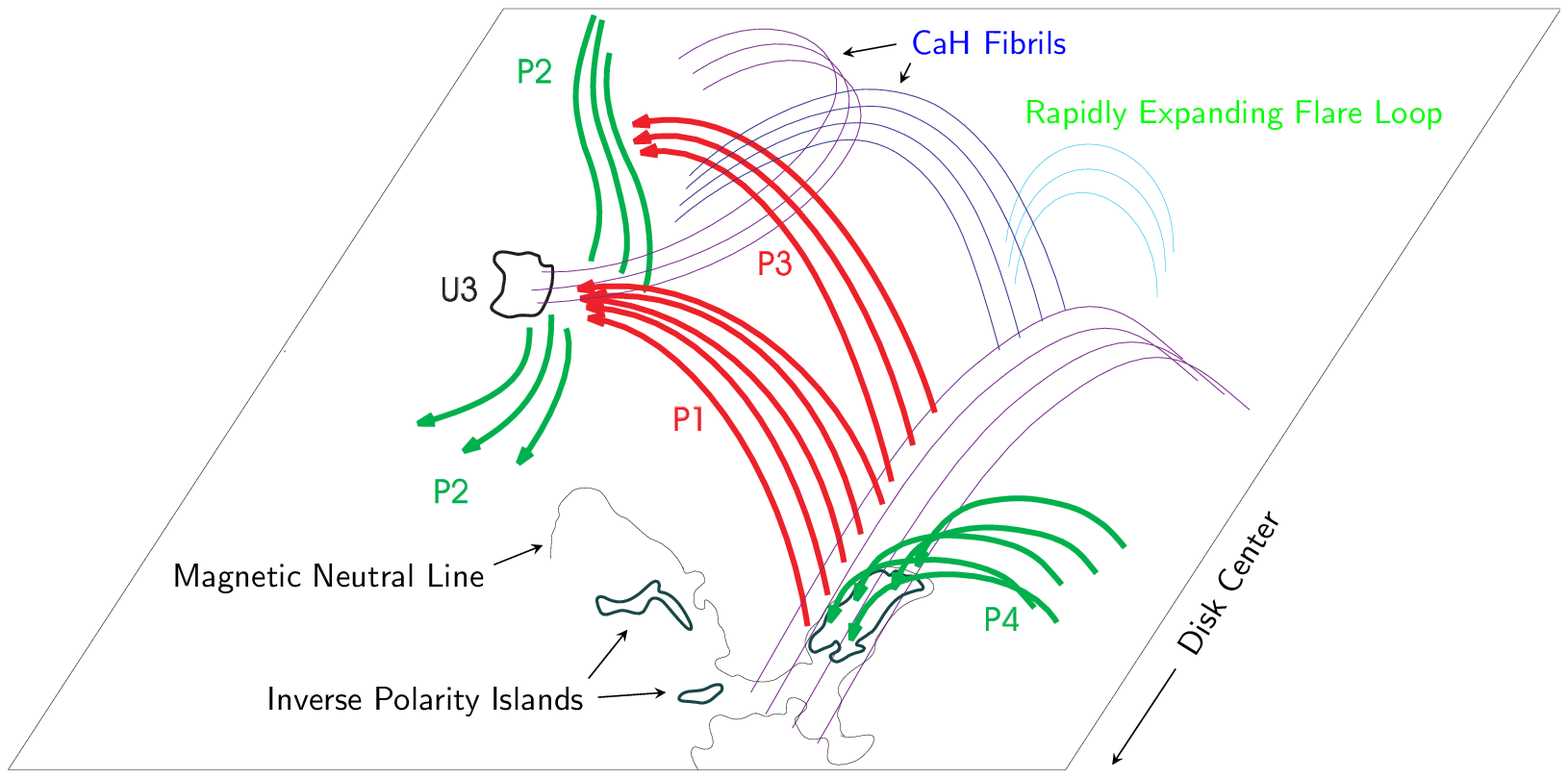} 

   \vspace*{-147mm}

   \caption{Schematic drawing of the disturbed penumbral region during
   the onset of the flare. Thick lines denote photospheric penumbral
   fibrils (colours of the different branches are identical with those
   in Fig.~\ref{fig2}; arrows delineate the magnetic field direction);
   thin coloured lines denote chromospheric \ion{Ca}{ii}\,$H$ fibrils;
   black contours mark the magnetic neutral line (thin) and the small
   umbra U3 (thick). The islands of inverse $B_z$ are displayed as
   thick dark-green contours. The flare onset regions are located
   close to intersection of P1 and P2 (FR1) and at the connection area
   of P1 and P4 (FR2).}
   \label{fig22}
   \end{figure*}

The observed disturbed penumbral region of the trailing spot of the
bipolar active region NOAA~10904 shows an extremely complex magnetic
field and flow configuration in the photosphere and a hardly less complex
structure of \ion{Ca}{ii}\,$H$ fibrils, which represent the lower
chromosphere. Summarising the deductions made in the previous
sections, in Fig.~\ref{fig22} the atmospheric structure in the regions of
the two flare ribbons is sketched. One flare ribbon (FR1) is located
close to the small umbra U3, roughly parallel to the penumbral branch
P2. The other flare ribbon (FR2) is located above the endpoints of
penumbral branches P1 and P3 where they meet P4.

Flare ribbon FR2 (cf. Fig.~\ref{fig14}) includes an island of inverse
$B_z$, which seems to contain the footpoints of loop-like magnetic
fields. In close proximity to this island, a branch of penumbral
filaments (P1) with a polarity of the magnetic field and the Evershed
flow, which is opposite to the main polarity of the spot penumbra,
emerges from the subsurface layers.

The region around flare ribbon FR1 (cf. Fig.~\ref{fig17}) shows
several crossing branches of penumbral filaments, particularly at the
endpoints (DR2 and DR3, see Fig.~\ref{fig9}) of P1 and P3, which carry
the oppositely directed Evershed motions and in which the magnetic
fields are oppositely directed to the main penumbral field direction.

Both regions are overlaid by several crossing branches of fibrils
visible in \ion{Ca}{ii}\,$H$. Some of these branches show almost no
motions, only strong brightness enhancements during the flare, and
others show rapid expansion. Some of the \ion{Ca}{ii}\,$H$ fibrils are
found to modify their azimuthal direction during the flare, such that
a fibril that starts strongly skewed relative to the underlying
photospheric penumbral fibril ends up being more nearly parallel.

The obtained structure can be fitted into a global picture of the
flare eruption as developed in Falconer et al. (2000). The observed
region includes several islands of negative magnetic polarity in the
positive polarity region of NOAA~10904. According to Falconer et
al. (2000), the islands of inverse polarity are connected to the
adjacent regions by short loops leading to a closed core field
enveloping the neutral line around the islands (see their
Fig.~3). Close to the magnetic neutral lines, the fields are strongly
sheared, which is very often the case prior to a flare eruption
(Hagyard et al. 1984). A separatrix surface encloses the islands. The
footpoints of the separatrix surface enclose the entire region. In our
case, this intersection of the separatrix with the surface marks the
footpoints of the branches of \ion{Ca}{ii}\,$H$ fibrils, which are
connected to the interior of the disturbed region and can be easily
identified by a continuous brightness enhancement prior to and after
the flare eruption.

%\clearpage

%
%----------------------------------------------------------------
%

%
%----------------------------------------------------------------
%
%----------------------------------------------------------------
%

   \begin{figure*}
   \centering

   %\vspace*{35mm}

%   \parbox[b]{10cm}{
   \hspace*{10mm}\includegraphics[width=10cm]{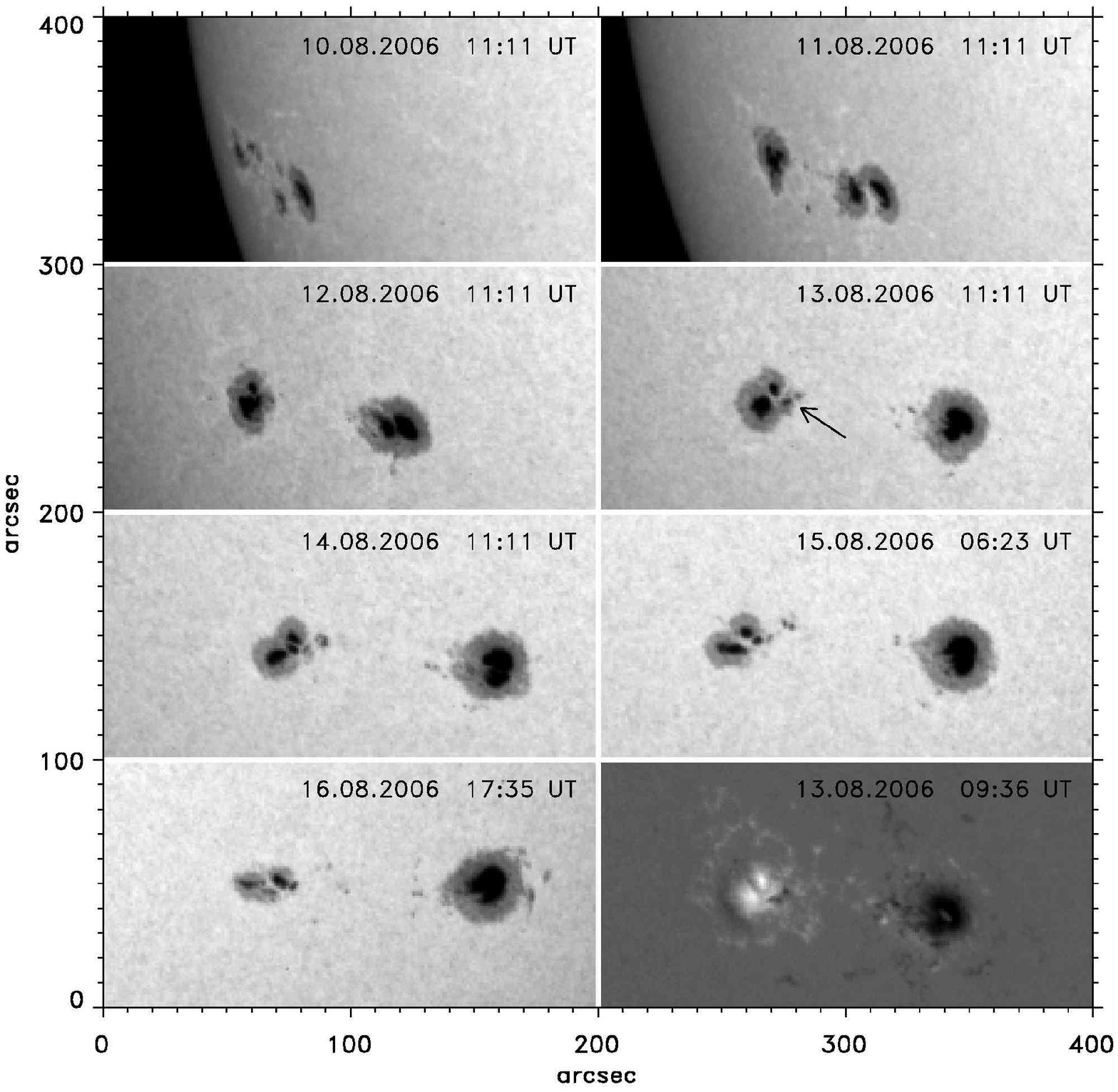}
%   }\hfill\parbox[b]{7cm}{

   %\vspace*{8mm}

   \caption{Evolution of active region
   NOAA~10904 as observed in white light by SOHO/MDI. The arrow point
   to the flare region and the lower right panel shows an MDI magnetic
   field map from August 13, 2006. The formation of the trailing
   sunspot remains unclear, but it decayed rapidly after August
   13, 2006.}
   \label{fig23}
%   }

   \vspace*{10mm}
   
   \end{figure*}
%
%----------------------------------------------------------------
%

Since the pre-flare structure of both the chromosphere and the
photosphere of the observed region does not differ much from the
post-flare structure and because the observed eruption is only a weak
flare (GOES magnitude B7.8) the global topology of the region does not
change during the flare. However, the high resolution of the analysed
data made it possible to study small-scale restructuring of the
atmosphere during a weak flare for the first time. Primarily, we found
a sizable increase in the islands of inverse polarity. This is
consistent with the long-term evolution of the spot, which can be
traced, e.g., on SOHO/MDI continuum images (see Fig.~\ref{fig23}). The
formation of the sunspot is not completely clear, since it appeared
for the first time on August~10, 2006 on the extreme eastern limb of
the Sun, but it seems that it was formed by a merger of two spots of
the same polarity. The two spots (represented by the two large umbrae
visible in Fig.~\ref{fig1}) achieved closest proximity around
August~12, 2006 and started to re-split (and decay) in the following
days until they were completely separated on August 16, 2006. The
observed crossing of penumbral branches of opposite polarity might
stem from the penetration of the penumbra of one spot into that of the
other one during the merger of the two spots. This penetration might
also drive the emergence of reversed polarity magnetic flux during the
flare eruption. The eruption even might be triggered by this flux
emergence. This interpretation is supported by several B-class events
in the considered active region having been registered by the GOES/SXI
instrument within several hours before and after the one discussed
here. However, this one with magnitude B7.8 was by far the strongest
one. The model of a restructuring of the intersecting penumbral
magnetic fields is, furthermore, supported by a single penumbral
filament aligned with the magnetic neutral line of one of the
inverse-polarity islands (see Fig.~\ref{fig21}) slowly decaying from
about 20\,min prior to the flare eruption on. Its last visible
remnants dissappear simultaneously with the onset of the flare.

Based on these results the actual triggering mechanism must also be
searched in the dynamics of the regions of the intersecting
penumbrae. However, the pre-flare dynamics of these regions only show
a few conspicuous features.  One of these features is a fast-moving
cloud visible in the \ion{Ca}{ii}\,$H$ line-core images (see
Fig.~\ref{fig20}). This cloud crosses the point of the flare onset (an
intersection of two branches of \ion{Ca}{ii}\,$H$ fibrils)
approximately one minute before the onset of the flare. The trajectory
of this cloud motion is co-spatial with the decaying penumbral
filament located at the neutral line of the magnetic island. We also
find a strongly sheared magnetic field configuration along this
neutral line. This conjuncture of events suggests a close connection
with the flare onset.

Other features are tiny dark structures crossing the bright heads of
dark-cored penumbral filaments. These hitherto unknown phenomena are
also located close to the previously mentioned magnetic neutral line
below the crossing of two branches of \ion{Ca}{ii}\,$H$ fibrils. While
the event takes place, the fibrils are already starting to increase
their intensity at a distance of about one arcsecond from the event.
The dark structures leave the penumbral filaments
unmodified. Therefore, we assume that they are some kind of dense
matter crossing the filaments in the upper photosphere. They are also
visible, although much less conspicuous, in the \ion{Ca}{ii}\,$H$
line-wing images. It is hard to decide whether these structures are
related to the triggering mechanism of the flare. More likely, they
are an early manifestation of the flare eruption.
  
The entire analysed disturbed penumbral region is connected with fast
motions. Most strikingly, we found strong photospheric Doppler shifts
that correspond to velocities of up to 7\,km\,s$^{-1}$. These motions
are most likely Evershed flows which achieve their maximum speeds
close to the onset points of the flare eruption. At these points,
convergence centres of the horizontal flows have also been detected
and the magnetic fields show conspicuous discontinuities. Therefore,
we speculate that the intersection of penumbrae may lead to a
contraction of their interlaced branches at certain points which
accelerates the Evershed flows. Several penumbral grains located in
close proximity to the convergence centres (see e.g. Fig.~\ref{fig8})
also move with unusually high speeds, which may or may not support our
hypothesis. Fast and even supersonic flows in confined penumbral
regions close to the magnetic neutral lines in $\delta$-spots were
already found by Mart\'\i nez Pillet et al. (1994) and Lites et
al. (2002), while evidence of supersonic Evershed flow in normal
penumbrae has been provided by Borrero et al. (2005) and Bellot Rubio
et al. (2007).

Fast motions were also detected in the chromosphere and upper
photosphere. Apart from a fast moving cloud visible in the
\ion{Ca}{ii}\,$H$ line-core images (see above), we found several
bright features flashing up in the \ion{Ca}{ii}\,$H$ line-wing
images. The simplest explanation for these flashes is a strong
blueshift of the \ion{Ca}{ii}\,$H$ line in these regions. The
line-wing images were obtained approximately 0.6\,\AA\ out of the line
centre. This value corresponds to blueshifts of 45\,km\,s$^{-1}$. We
may, therefore, assume velocities of several
$10^4$\,m\,s$^{-1}$. Beside these flashes, the only feature visible
in the \ion{Ca}{ii}\,$H$ line-wing and related to the flare, which
lasts longer than a few images, is the outgoing brightness front of a
rapidly expanding branch of loop-like \ion{Ca}{ii}\,$H$ fibrils. The
front, visible in the line-wing, might be assumed to correspond to
the footpoints of the loop-like structures. There the chromospheric
magnetic field lines should be mainly vertical which is consistent
with fast upflows.

%
%----------------------------------------------------------------
%
% Fig10online available electronically only

\onlfig{24}{
\begin{figure*}%f2

%\vspace*{3mm}

\hspace*{10mm}\includegraphics[width=15cm]{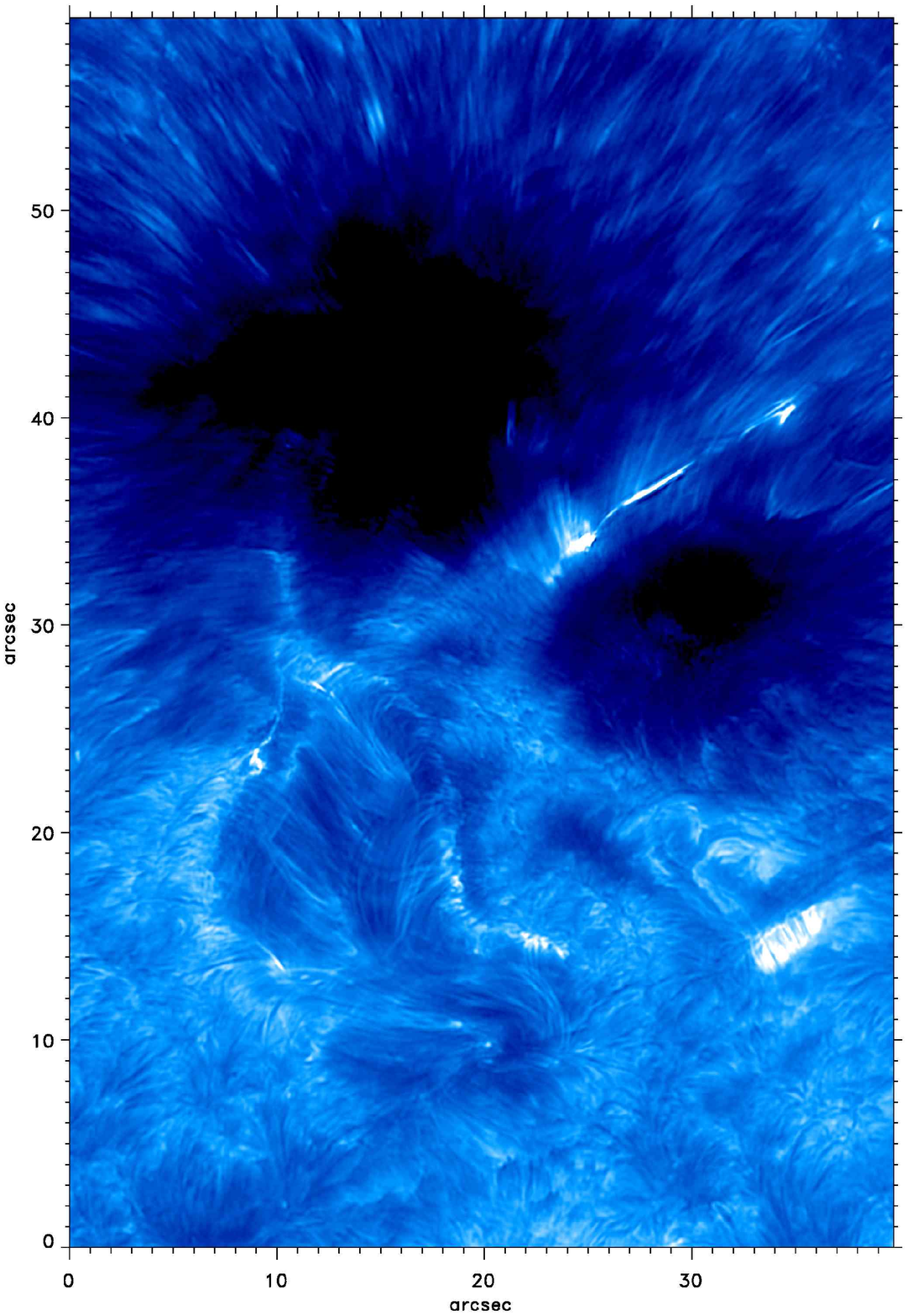}

%\vspace*{10mm}

\caption {Full field of viw of the \ion{Ca}{ii}\,$H$ observations. The
          image shows the sunspot and the flaring region at 9:58\,UT
          which is about one hour after the flare eruption.}
\label{fig10online}
\end{figure*}
}

%
%----------------------------------------------------------------
%

\begin{acknowledgements}
       The authors are grateful to an anonymous referee for helpful
       comments. Part of this work was supported by the WCU grant
       (No. R31-10016) from the Korean ministry of Education, Science
       and Technology. The Swedish Solar Telescope (SST) is operated
       on the island of La Palma by the Institute for Solar Physics of
       the Royal Swedish Academy of Sciences in the Spanish
       Observatorio del Roque de los Muchachos of the Instituto de
       Astrif\'\i sica de Canarias. We thank the telescope stuff,
       Helena Uthas, Rolf Kever and Michiel van Noort, for their kind
       support with the SST.
\end{acknowledgements}

%
%----------------------------------------------------------------
%

%
%
% Online Material
%_____________________________________________________________
%        Online appendices have to be placed at the end, after
%                                        \end{thebibliography}
%-------------------------------------------------------------

\end{document}